\documentclass[prd,preprint,tightenlines,aps,nofootinbib,preprintnumbers,floatfix]{revtex4}
\usepackage{amsmath,amssymb}
\usepackage{epsfig}
\usepackage{graphicx}
\usepackage{subfigure}

\def\lsim{\mathrel{\hbox{\rlap{\hbox{\lower4pt\hbox{$\sim$}}}\hbox{$<$}}}}
\def\gsim{\mathrel{\hbox{\rlap{\hbox{\lower4pt\hbox{$\sim$}}}\hbox{$>$}}}}
\def\lesssim{\mathrel{\hbox{\rlap{\hbox{\lower4pt\hbox{$\sim$}}}\hbox{$<$}}}}
\def\gtrsim{\mathrel{\hbox{\rlap{\hbox{\lower4pt\hbox{$\sim$}}}\hbox{$>$}}}}
\def\be{\begin{equation}}
\def\ee{\end{equation}}
\def\bea{\begin{eqnarray}}
\def\eea{\end{eqnarray}}

\def\Mp{M_{\mathrm{Pl}}}
\def\tp{t_{\mathrm{Pl}}}

\begin{document}
\title{Cosmic Super-Strings and Kaluza-Klein Modes}
\date{\today}
\author{Jean-Fran\c{c}ois Dufaux}
\affiliation{APC, Univ. Paris Diderot, CNRS/IN2P3, CEA/Irfu, Obs. de Paris, Sorbonne Paris Cit\'e, France}

\begin{abstract}
Cosmic super-strings interact generically with a tower of relatively light and/or strongly coupled Kaluza-Klein (KK) modes associated with the geometry of the internal space. In this paper, we study the production of spin-$2$ KK particles by cusps on loops of cosmic F- and D-strings. We consider cosmic super-strings localized either at the bottom of a warped throat or in a flat internal space with large volume. The total energy emitted by cusps in KK modes is comparable in both cases, although the number of produced KK modes may differ significantly. We then show that KK emission is constrained by the photo-dissociation of light elements and by observations of the diffuse gamma ray background. We show that this rules out regions of the parameter space of cosmic super-strings that are complementary to the regions that can be probed by current and upcoming gravitational wave experiments. KK modes are also expected to play an important role in the friction-dominated epoch of cosmic super-string evolution.
\end{abstract}

\maketitle


\section{Introduction}

Cosmic strings are linear concentrations of energy with a cosmological size, which appear in several high-energy physics scenarios and may have a variety of cosmological and astrophysical consequences~\cite{VS, HK}. Field theory cosmic strings have been abundantly studied since the 70's~\cite{kibble, NO}. The subject has been recently regenerated by the progressive realization~\cite{witten, majumdar, tye, dvalenkin, CMP} that cosmic strings occur also as fundamental objects of string theory, see \cite{pol, daviskibble, majumdar2, sakel, copeland} for reviews. These so-called cosmic super-strings are expected to be produced in particular at the end of brane inflation~\cite{DvaliTye, tye, dvalenkin} or at Hagedorn phase 
transitions~\cite{englert, majumdar} after inflation~\cite{pol, chen}. This opens the opportunity that they could lead to observational signatures of string theory, if these signatures can be distinguished to some degree from the ones of traditional field theory cosmic strings. Properties of cosmic super-strings that may help this distinction, although they appear also in some field theory models, include the possibilities of reconnection probabilities smaller than unity and of strings with different charges and junctions between them. In this paper, we consider another potentially specific aspect of cosmic super-strings: their interactions with massive Kaluza-Klein (KK) modes associated with the compact extra dimensions. We will study in particular the production of KK particles by cosmic super-strings~\cite{letter}.

\begin{figure}[htb]
\begin{center}
\includegraphics[width=7cm]{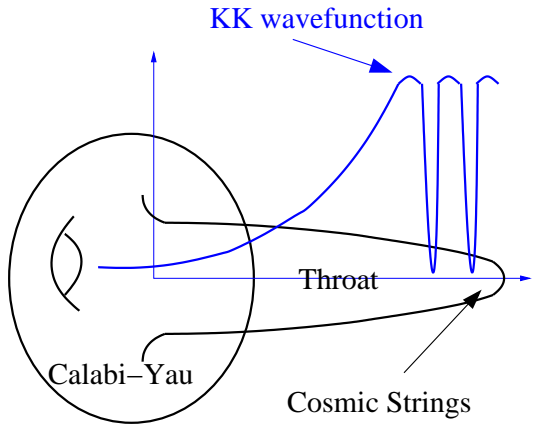}
\end{center}
\vspace*{-5mm}
\caption{Sketch of a compact throat with cosmic super-strings and KK modes localized at the bottom.}
\label{throatCY}
\end{figure}

The extra dimensions play a crucial role for cosmic super-strings, being responsible for making their tension small enough compared to the Planck scale, as required by observations. This can be achieved in two different ways~\cite{pol}. The first possibility is to have cosmic strings localized at the bottom of a warped ten-dimensional geometry, called throat, where the effective four-dimensional scales are "redshifted" by a warp factor that depends on the position in the internal 
space~\cite{RS}. This occurs in particular in models of warped brane inflation~\cite{KKLMMT} with stabilized 
moduli~\cite{GKP,KKLT}, which currently represent one of the most popular class of inflationary models in string theory. In these models, the effective string tension $\mu$ in units of the four-dimensional Newton constant $G$ is exponentially small because of the value $e^{A_b} \ll 1$ of the warp factor at the bottom of the throat, $G \mu \propto e^{2 A_b}$. In that case, the lowest-lying KK modes are lighter than the string scale $\sqrt{\mu}$ because their masses are also suppressed by the warp factor $e^{A_b}$. Furthermore, their wave-function in the extra-dimensions is localized at the bottom of the throat, where the cosmic strings are located (see Fig.~\ref{throatCY}). As a result, their coupling to the cosmic strings is enhanced by the factor $e^{-A_b} \gg 1$ with respect to the gravitational coupling. Thus these KK modes are relatively strongly coupled to the cosmic super-strings. The second possibility is to have a flat internal space with a large six-dimensional volume~\cite{ADD, balasub}, $V_6 \gg \alpha'^3$ where $\sqrt{\alpha'}$ is the fundamental string length, so that $G \mu \propto \alpha'^3 / V_6$. In that case, the KK modes couple only gravitationally to the cosmic strings, but there is now a dense spectrum of modes with masses much smaller than $\sqrt{\mu}$. Thus a lot of these KK modes can then be produced. Whether KK modes are strongly coupled or very light, they can be efficiently produced by cosmic super-strings in both cases.

The tension of cosmic super-strings is highly model-dependent, depending in particular on the values 
of $V_6$ and $e^{A_b}$ and on the energy scale at which the cosmic strings form. For cosmic super-strings produced at the end of brane inflation, the tension depends on the inflationary energy scale, which in turn is constrained by the Cosmic Microwave Background (CMB). In models of brane inflation where the CMB anisotropies are generated only by the quantum fluctuations of a single inflaton field, the string tension is expected to be in the approximate ranges
$10^{-12} \lesssim G \mu \lesssim 10^{-7}$ for a flat internal space with large volume~\cite{tye} and 
$10^{-14} \lesssim G \mu \lesssim 10^{-7}$ for a warped throat~\cite{shandera}. The gravitational effects of such cosmic strings may be observable in the near future, in particular with upcoming gravitational wave (GW) experiments~\cite{DV, DVsuper, siemens, hogan, elisa}. In general, however, cosmic super-strings may have a much smaller tension. This occurs in particular when they are produced in models of brane inflation at lower energy scales. Indeed these models have now been generalized in a number of ways and the CMB anisotropies may be generated from quantum fluctuations of other fields, like e.g. in the curvaton~\cite{curvaton} and modulated reheating~\cite{modulated} scenarios. For instance, it was shown in 
\cite{lev} that the conformal-coupling problem of the original model of warped brane inflation~\cite{KKLMMT} can be made harmless if inflation occurs at low energy scales, as low as the TeV scale. This leads to cosmic super-strings with tensions as small as $G \mu \sim 10^{-34}$. The tension of cosmic super-strings is also relatively small when they form at Hagedorn phase transitions~\cite{englert, majumdar} after inflation. For instance, if one throat is present in the internal space, then it is maybe more typical to have several of them, which also offers more opportunities to naturally generate several hierarchies. In these multi-throat scenarios, cosmic super-strings are produced in long throats that are heated above their local Hagedorn temperature after inflation~\cite{pol, chen}. These can be much lighter than the inflationary energy scale. Because the gravitational effects of such light cosmic strings are much weaker, it is crucial to look for their other possible signatures, in particular the production of (massive) particles.

There are at least three general motivations to look for particle production by cosmic strings (besides gravitational waves), depending on the model and on the nature of these particles. First, the particles produced by strings may have useful cosmological consequences, e.g. for baryogenesis~\cite{baryo}, ultra-high energy cosmic rays~\cite{cosmicray, vach, sabancilar}, or the non-thermal production of dark matter~\cite{darkmatter}. Secondly, these particles may lead instead to cosmological problems, and therefore to constraints on the cosmic string parameters, see e.g.~\cite{DVdil, PS, ModShort, ModLong, instant, flatdir, flatdir2}. Finally, particle production may affect both the dissipation of small scale structure on the long strings and the rate at which the loops shrink, hence the predictions for other cosmic string signatures. 

The production of massive particles by field theory cosmic strings has been studied by several authors. Ref.~\cite{sred} considered a scalar field with a quartic interaction with cosmic strings and studied its production from cusps - highly boosted regions where the string doubles back on itself. Refs.~\cite{vach,sabancilar} considered instead trilinear interactions, which make particle production more efficient and may occur when a scalar field condenses on the string or for strongly coupled moduli fields. The production of gravitationally coupled moduli fields was studied in \cite{DVdil, PS}. Besides these perturbative processes of particle production, cosmic strings also emit quanta of their constituent fields when string segments annihilate. This occurs in particular at cusps on usual Abelian-Higgs cosmic strings, as first suggested in \cite{robert} and revisited in detail in \cite{BPO}. Massive particles may also be copiously produced in more complicated models, e.g. when the cosmic strings are super-conducting~\cite{superconducting}.   

Cosmic super-strings, on the other hand, are more likely to be stable if they are somehow decoupled from other degrees of freedom, in particular from the Standard Model fields~\cite{CMP}. The kind of particles they can emit is therefore more constrained~\footnote{The production of massless axionic radiation by cosmic super-strings has been discussed 
in~\cite{RR, RR2}.}. However, they always couple to the ten-dimensional metric. From the four-dimensional point of view, 
this generically includes, beyond the ubiquitous four-dimensional massless graviton, a tower of relatively light KK modes associated with the geometry of the extra dimensions. 

In a recent letter~\cite{letter}, we studied the production of KK modes by cosmic super-strings in a warped throat, showing that it is contrained by the photo-dissociation of the light elements produced by Big Bang Nucleosynthesis. These constraints rule out regions of the parameter space of cosmic super-strings that are complemetary to the ones that can be probed by GW experiments. In the present paper, we give a full computation of these results. In addition to~\cite{letter}, we also consider cosmic super-strings located in a flat internal space with large volume, we study the constraints on KK emission coming from observations of the diffuse gamma ray background, and we discuss the effects of KK modes on the early, friction-dominated stage of cosmic super-string evolution. 

The rest of the paper is organized as follows. In Section~\ref{coupling}, we consider the spectrum of KK modes and their coupling to cosmic F- and D-strings. We then study the emission of KK modes by cosmic string loops in 
Section~\ref{emission}. Section~\ref{distribution} is dedicated to the consequences of KK emission for the loop number density and the effect of the friction-dominated epoch for cosmic super-strings. We then address cosmological consequences of KK emission and constraints on the cosmic string parameters in Section~\ref{constraints}. We conclude in Section~\ref{conclu} with a summary of our results and directions for future work. An Appendix is dedicated to computational aspects of the emission of massive particles by cusps.

\section{Kaluza-Klein Coupling to Cosmic Super-Strings}
\label{coupling}

In this section, we discuss the spectrum of KK modes and their coupling to cosmic F- and D-strings. As discussed in the introduction, cosmic super-strings may live either at the bottom of a warped ten-dimensional geometry or in a flat internal space with large volume. We consider both cases in turn.

\subsection{Warped Throat}
\label{throat}

Consider a Nambu-Goto string in a ten-dimensional space-time with metric $\tilde{g}_{A B}$ ($A, B = 0, ..., 9$). 
The system is described by the action~\footnote{We only consider the interaction of cosmic super-strings with the ten-dimensional metric, because this will be relevant one for the KK modes that we will consider.}
\be
\label{action}
S = \frac{M_{10}^8}{2}\,\int d^{10}x\,\sqrt{-\tilde{g}}\,\left(\mathcal{R} + 2 \mathcal{L}\right) - 
\frac{g_s^{-\chi}}{2\pi\,\alpha'}\,\int d\tau d\sigma\,\sqrt{-\tilde{\gamma}} 
\ee 
where $M_{10}$ is the ten-dimensional Planck mass, $\mathcal{R}$ is the ten-dimensional Ricci scalar, $\mathcal{L}$ denotes the Lagrangian of all the bulk matter fields (scalar fields, forms, brane localized terms, ...) and $\tilde{\gamma}$ is the determinant of the two-dimensional metric induced on the string worldsheet
\be
\label{induced}
\tilde{\gamma}_{\alpha \beta} = \tilde{g}_{AB}\,\partial_{\alpha} X^A\,\partial_{\beta} X^B
\ee 
where $\alpha, \beta = 0, 1$ and the embedding of the worldsheet in the ten-dimensional spacetime is parametrized as 
$x^A = X^A(\tau, \sigma)$. In the second term of Eq.~(\ref{action}), $1 / (2 \pi\,\alpha')$ and $g_s$ are the fundamental string tension and string coupling, respectively. For F-strings, we have $\chi = 0$, while for D-strings the tension is increased by $1/g_s$, i.e. $\chi = 1$ in that case. In the following, we will assume that only one kind of string is present or dominate the loop number density. 

In a warped throat, the ten-dimensional metric is a warped product of our four-dimensional spacetime with a six-dimensional internal space
\be
\label{metric}
ds^2 = \tilde{g}_{AB}\,dx^A\,dx^B = e^{2 A(\bar{y})}\,g_{\mu \nu}\,dx^{\mu}\,dx^{\nu} + \hat{g}_{ab}(\bar{y})\,dy^a\,dy^b
\ee 
where $\bar{y}$ denotes the coordinates in the internal space, $\mu, \nu = 0, ..., 3$ and $a, b = 4, ..., 9$. The background solution with four-dimensional Poincare invariance is given by $g_{\mu \nu} = \eta_{\mu \nu}$. The most studied example of throat geometry is based on the Klebanov-Strassler solution of type II B supergravity with fluxes~\cite{KS}. In most part of the throat, the metric can be approximated by the direct product of five-dimensional anti-de sitter spacetime with a five-dimensional manifold called $T^{1,1}$
\be
\label{dsapprox}
ds^2 \approx e^{-2 y / \tilde{R}}\,\eta_{\mu \nu}\,dx^{\mu}\,dx^{\nu} + dy^2 + \tilde{R}^2\,d\Omega_5^2
\ee
where $\tilde{R}$ is the radius of curvature of $AdS_5$ and $T^{1,1}$, $y$ is a radial coordinate of $AdS_5$ and 
$d\Omega_5^2$ is the metric of $T^{1,1}$. The ratio $\tilde{R} / \sqrt{\alpha'}$ depends on an integer number of flux 
units $M$,
\be
\label{gsM}
\frac{\tilde{R}}{\sqrt{\alpha'}} \approx \sqrt{g_s\,M} \, ,
\ee
and should be larger than unity in order to trust the supergravity approximation. It will be convenient to define~\footnote{The radius of curvature $\tilde{R}$ was denoted $R$ in \cite{letter}. Here we use $R$ to denote $2 \pi \, \tilde{R}$, because the relevant equations are then formally identical to the equations in the case of a flat internal space with large volume that will be discussed in the next sub-section.}
\be
\label{defR}
R = \frac{\tilde{R}}{2\pi}
\ee
in order to simplify notations in the following. 

The approximation (\ref{dsapprox}) for the metric is only valid for some range of the radial coordinate, $0 < y < y_b$. At $y = y_b$, the $T^{1,1}$ evolves into a round three-sphere of finite radius with the two remaining angular dimensions shrinking to zero size. This corresponds to the bottom of the throat, where the warp factor takes its minimal value $e^{A_b} \ll 1$. At $y = 0$, the Klebanov-Strassler solution is glued to a compact Calabi-Yau where the warp factor is approximately constant, $e^{A} \approx 1$. The total volume of the internal space is dominated by the volume $V_6 > \tilde{R}^6$ of the compact Calabi-Yau. The four-dimensional Newton constant $G$ is then given by
\be
\label{GN}
\frac{1}{8\pi\,G} = M_{10}^8\,V_6 = \frac{2\,V_6}{\left(2\pi\right)^7\,g_s^2\,\alpha'^4}
\ee
where in the second equality we have used the definition of the ten-dimensional Planck mass in type II B string theory.

The cosmic strings are localized at the bottom of the throat, because the effective potential for their position in the internal space is minimum there~\cite{CMP}. As emphasized in \cite{psuper}, the position of the cosmic strings in the internal space corresponds to worldsheet moduli that are not protected by any symmetry, and should then be fixed at the minimum of their effective potential (except potentially in the very early universe) at the bottom of the 
throat~\footnote{The strings could still move freely on the three-sphere at the bottom of the throat if its isometries were exact. However, these isometries are broken by the Calabi-Yau glued at $y=0$, which provides an effective potential for the position of the strings on the three-sphere~\cite{aharony}.}: $X^A(\tau, \sigma) = \mathrm{constant}$ at the classical level for $A = 4, ..., 9$. Defining
\be
\label{gamma}
\gamma_{\alpha \beta} = \eta_{\mu \nu}\,\partial_\alpha X^\mu\,\partial_\beta X^\nu \, ,
\ee  
the second term in Eq.~(\ref{action}) then reduces to the standard form
\be
\label{NGS}
- \mu\,\int d\tau d\sigma\,\sqrt{-\gamma}
\ee
where
\be
\label{muwarp}
\mu = \frac{M_s^2}{2 \pi\,g_s^\chi} = \frac{e^{2 A_b}}{2\pi\,\alpha'\,g_s^\chi}
\ee
is the effective string tension perceived in four dimensions and $M_s = e^{A_b} / \sqrt{\alpha'}$ is the local string mass scale at the bottom of the throat.  

We now consider the KK modes associated with the warped geometry. We will focus on the modes that correspond to spin-$2$ fields in four dimensions, because these are the simplest and most generic ones. These modes are always present and the massless one corresponds to the usual four-dimensional graviton. Basic properties of spin-$2$ KK modes in warped throats are well understood, see e.g.~\cite{KKrelics} and references therein. Their wavefunction in the internal space is the same as the one of a test scalar field in the same geometry, independently of the matter Lagrangian $\mathcal{L}$ in Eq.~(\ref{action}). The KK modes corresponding to scalar and vector fields in four dimensions are more model-dependent, depending in particular on the symmetries and the mechanisms of moduli stabilization. The spin-$2$ KK modes correspond to a metric perturbation that is transverse and traceless in four dimensions. They can be described by the metric (\ref{metric}) with
\be
g_{\mu \nu} = \eta_{\mu \nu} + \sum_{\bar{n}} \Phi_{\bar{n}}(\bar{y})\,h^{\bar{n}}_{\mu \nu}(x)
\ee
where $\bar{n}$ denotes collectively the mode numbers associated to each KK modes. For instance, in the approximation 
(\ref{dsapprox}), $\bar{n}$ is made of the mode numbers associated with the Laplace operator on $T^{1,1}$ plus one mode number associated with the radial coordinate $y$. The $h^{\bar{n}}_{\mu \nu}(x)$ are four-dimensional spin-$2$ fields of mass $m_{\bar{n}}$, satisfying
\be
\left(\Box + m_{\bar{n}}^2\right)\,h^{\bar{n}}_{\mu \nu}(x) = 0 \hspace*{0.5cm} , \hspace*{0.5cm} 
\partial^\mu h^{\bar{n}}_{\mu \nu} = h^{\bar{n}\,\lambda}_\lambda = 0 
\ee
for the free fields. Here and in the following, the four-dimensional Greek indices are raised and lowered with the Minkowski metric $\eta_{\mu \nu}$, and $\Box = \partial^\lambda \partial_\lambda$ is the four-dimensional d'Alembertian. 

The properties of the KK modes in the warped geometry are reminiscent to the Randall-Sundrum model~\cite{RS}. The massless mode $m=0$, corresponding to the usual four-dimensional graviton, has a constant wavefunction in the internal space. By contrast, the massive modes are strongly localized at the bottom of the throat (see Fig.~\ref{throatCY}) and their masses are quantized as
\be
\label{mwarp}
m_{\bar{n}} = c_{\bar{n}}\,\frac{e^{A_b}}{R} 
\ee
where $c_{\bar{n}}$ is a numerical coefficient that depends on the mode numbers, with $c_{\bar{n}} \sim 1$ for the 
lower-lying KK modes. These massive modes are generic in a warped throat. Other KK modes of mass $m \sim V_6^{-1/6}$ may be localized instead in the compact Calabi-Yau at $y \sim 0$, see e.g.~\cite{box}. However, for sufficient warping, 
$e^{A_b} / R \ll V_6^{-1/6}$, these latter modes are much heavier than the former ones and their coupling to the cosmic super-strings is exponentially smaller. We will therefore not consider them here. The case of large volume $V_6$ will be discussed in the next sub-section. Another possibility is that the cosmic strings are located at the bottom of a warped throat but that other, longer throats with more warping are also present. The lightest KK modes are then localized at the bottom of the longest throat. Even in that case, the KK modes localized in the same throat as the cosmic strings are still present as resonant states in the KK spectrum, see e.g.~\cite{langfelder}. These are the modes we consider because the modes localized in other throats are practically decoupled from the cosmic strings. 

With the ortho-normalization condition
\be
\frac{M_{10}}{8}\,\int d^6\bar{y}\,\sqrt{\hat{g}(\bar{y})}\,e^{2A(\bar{y})}\,
\Phi_{\bar{n}}(\bar{y})\,\Phi_{\bar{n}'}(\bar{y}) = \delta_{\bar{n} \bar{n}'}
\ee
for the wavefunction in the internal space, the ten-dimensional action in (\ref{action}) expanded to quadratic order in 
$h^{\bar{n}}_{\mu \nu}$ reduces to the canonical action for four-dimensional spin-$2$ fields after integration over the internal space, see \cite{KKrelics} for more details. The constant wavefunction of the massless mode then reads
\be
\label{phi0}
\Phi_{\bar{0}}(\bar{y}) = 2\,\sqrt{8\pi\,G} \, ,
\ee
where Eq.~(\ref{GN}) has been used. The wavefunction of the massive modes has an enhanced amplitude at the bottom of the throat
\be
\label{phim}
\Phi_{\bar{n}}(\bar{y}_b) \simeq 2\,\sqrt{16\pi\,G}\,\left(\frac{V_6}{\tilde{R}^6}\right)^{1/2}\,e^{-A_b} 
\hspace*{0.5cm} \mbox{ for } m_{\bar{n}} \neq 0
\ee
where $\bar{y} = \bar{y}_b$ denotes the bottom of the throat.

We can now determine the coupling of the KK modes to the cosmic string. This is obtained by expanding the induced metric 
$\tilde{\gamma}_{\alpha \beta}$ in Eq.~(\ref{induced}) to first order in $h^{\bar{n}}_{\mu \nu}$ in the second term of 
(\ref{action}). This leads to the interaction term
\be
\label{int1}
- \, \frac{\mu}{2}\,\Phi_{\bar{n}}(\bar{y}_b)\,\int d\tau d\sigma\,\sqrt{-\gamma}\,\gamma^{\alpha \beta}\,
\partial_{\alpha} X^\mu \, \partial_{\beta} X^\nu\,h^{\bar{n}}_{\mu \nu}
\ee
for each KK mode, where we have used Eqs.~(\ref{gamma}) and (\ref{muwarp}), and $\Phi_{\bar{n}}(\bar{y})$ is evaluated at the position of the string, i.e. at the bottom of the throat $\bar{y} = \bar{y}_b$. It will be convenient to consider the coupling in units of the effective four-dimensional string tension $\mu$, by defining
\be
\label{deflam}
\lambda_{\bar{n}} = \frac{\sqrt{\mu}}{2}\,\Phi_{\bar{n}}(\bar{y}_b) \, .
\ee 
Using Eqs.~(\ref{muwarp}), (\ref{phi0}), (\ref{phim}) and (\ref{GN}), this gives
\be
\label{lambdawarp}
\lambda_{\bar{n}} \, =
\left\{
\begin{array}{l}
\sqrt{8\pi\,G\,\mu} \hspace*{1.55cm} \mbox{for } \, m_{\bar{n}} = 0 \\
g_s^{1 - \chi/2}\,\left(\frac{\sqrt{\alpha'}}{R}\right)^3 
\hspace*{0.5cm} \mbox{for } \, m_{\bar{n}} \neq 0 
\end{array}
\right.
\ee
where $R$ is defined in Eq.~(\ref{defR}). Note that the massive modes may couple much more strongly to the string than the massless mode, in particular for small values of the string tension. The interaction term (\ref{int1}) may now be 
re-written as
\be
\label{int2}
- \, \frac{\lambda_{\bar{n}}}{\sqrt{\mu}}\,\int d^4x\,h^{\bar{n}}_{\mu \nu}\,T^{\mu \nu}
\ee
where
\be
\label{Tmunu}
T^{\mu \nu} = \mu\,\int d\tau d\sigma\,\sqrt{-\gamma}\,\gamma^{\alpha \beta}\,\partial_{\alpha} X^\mu \, 
\partial_{\beta} X^\nu\, \delta^{(4)}\left(x^\lambda - X^\lambda(\tau,\sigma)\right)
\ee
is the four-dimensional energy-momentum tensor of the string.

\subsection{Flat Internal Space with Large Volume}
\label{flat}

As discussed in the Introduction, cosmic super-strings may also be located in a flat internal space with large volume. In that case, the four-dimensional string tension $\mu$ is comparable to the ten-dimensional one
\be
\label{muflat}
\mu = \frac{M_s^2}{2 \pi\,g_s^\chi} = \frac{1}{2\pi\,\alpha'\,g_s^\chi} \, .
\ee 
Thus, contrary to Eq.~(\ref{muwarp}), $\mu$ is not anymore reduced by a warp factor. The observational constraint 
$G \mu \ll 1$ then requires $G \ll \alpha'$, which can be satisfied if the volume of the internal space is sufficiently large. We will consider a simple scenario of toroidal compactification where $d$ extra dimension(s) have the same large radius $R \gg \sqrt{\alpha'}$, while the remaining $6-d$ extra dimensions have the minimal radius $\sqrt{\alpha'}$. The volume of the internal space is then given by
\be
V_6 = \left(2\pi\,R\right)^d\,\left(2\pi\,\sqrt{\alpha'}\right)^{(6-d)}\, .
\ee
Using Eqs.~(\ref{GN}) and (\ref{muflat}), this gives
\be
\label{Gmuflat}
G\,\mu = \frac{g_s^{2 - \chi}}{16 \pi}\,\left(\frac{\sqrt{\alpha'}}{R}\right)^d
\ee 
for the string tension in units of the four-dimensional Newton constant $G$.

The properties of the KK modes in this scenario are the same as in \cite{ADD}. For the massive modes, the normalized wavefunctions in the $d$ large extra dimensions $\bar{y} = (y_1, ..., y_d)$ are given by
\be
\label{phiflat}
\Phi_{\bar{n}}(\bar{y}) = 2\,\sqrt{16 \pi\,G}\,\cos\left(\frac{\bar{n} . \bar{y}}{R}\right)  
\hspace*{0.5cm} \mbox{ for } m_{\bar{n}} \neq 0
\ee
where $\bar{n} = (n_1, ..., n_d)$, while the zero mode is the same as in Eq.~(\ref{phi0}). The $n_i$ are the mode numbers associated with each large dimension and they take integer values. The masses of the KK modes are quantized as
\be
\label{mflat}
m_{\bar{n}} = \frac{c_{\bar{n}}}{R} \hspace*{0.5cm} \mbox{with} \hspace*{0.5cm} 
c_{\bar{n}} = |\bar{n}| = \sqrt{n_1^2 + ... + n_d^2}
\ee
instead of (\ref{mwarp}). Note that several KK modes with different mode numbers have the same mass, i.e. some mass levels are degenerate. Isometry breaking may lift this degeneracy, but if this effect is perturbative it should not modify the total number of KK modes that are lighter than the string scale.  

For a flat internal space, the massive modes (\ref{phiflat}) have now a constant amplitude in the extra dimensions, given by $2\, \sqrt{16\pi\,G}$. The modes still interact as in (\ref{int2}) with the cosmic string, but now the coupling is suppressed by $G$ for all the modes
\be
\label{lambdaflat}
\lambda_{\bar{n}} \, =
\left\{
\begin{array}{l}
\sqrt{8\pi\,G\,\mu} \hspace*{4.15cm} \mbox{for } \, m_{\bar{n}} = 0 \\
\sqrt{16\pi\,G\,\mu} = g_s^{1 - \chi/2} \,\left(\frac{\sqrt{\alpha'}}{R}\right)^{d/2}
\hspace*{0.5cm} \mbox{for } \, m_{\bar{n}} \neq 0 
\end{array}
\right.
\ee
where we have used Eq.~(\ref{Gmuflat}). On the other hand, the lower-lying KK modes are now much lighter than the string scale
\be
\label{moverM}
\frac{m_{\bar{n}}}{M_s} = c_{\bar{n}}\,\frac{\sqrt{\alpha'}}{R} \, .
\ee

Note that, for $d = 6$, the coupling (\ref{lambdaflat}) in terms of the compactification radius $R$ is formally identical to the coupling (\ref{lambdawarp}) of the massive modes in a warped throat, where in the latter case $R$ is defined in Eq.~(\ref{defR}) in terms of the curvature radius of the throat $\tilde{R}$. Similarly, Eq.~(\ref{moverM}) applies also to the case with warping, see Eqs.~(\ref{muwarp}) and (\ref{mwarp}). The main difference is that, in a warped throat, a large hierarchy between $\sqrt{\alpha'}$ and $R$ is not required. By contrast, in the case of large and flat internal space, 
$\sqrt{\alpha'} \ll R$ is required by the observational constraint $G\,\mu \ll 1$. This decreases the coupling of each KK mode to the cosmic strings, but increases the number of KK modes that are lighter than the string scale.

\section{Kaluza-Klein Emission by Cusps}
\label{emission}

We now study the production of KK modes by a cosmic super-string loop. More details about some parts of the calculation are given in the Appendix.

From the interaction (\ref{int2}), the energy emitted in spin-$2$ KK modes with mode numbers $\bar{n}$ and mass 
$m_{\bar{n}} > 0$ by a cosmic string loop can be calculated as
\be
\label{Em}
E_{\bar{n}} = \frac{\lambda_{\bar{n}}^2}{2\,\mu}\,\int \frac{d^3\mathbf{k}}{(2 \pi)^3}\,
\left(T^{\mu\nu}(\omega_k, \mathbf{k})\,T_{\mu\nu}^{*}(\omega_k, \mathbf{k}) - 
\frac{1}{3}\,|T^{\lambda}_{\lambda}(\omega_k, \mathbf{k})|^2\right) \hspace*{1cm} (m_{\bar{n}} \neq 0) 
\ee
where
\be
\label{FT}
T_{\mu\nu}(\omega_k, \mathbf{k}) = \int d^4x\,T_{\mu\nu}(t, \mathbf{x})\,e^{i k_\lambda x^\lambda}
\ee
is the Fourier transform of the loop energy-momentum tensor with respect to space and time, and we have introduced the 
$4$-vector
\be
\label{4k}
k^\lambda = (\omega_k, \mathbf{k}) \hspace*{0.5cm} , \hspace*{0.5cm} \omega_k = \sqrt{k^2 + m_{\bar{n}}^2} 
\ee
where $k = |\mathbf{k}|$ is the norm of the $3$-vector $\mathbf{k}$. The massive spin-$2$ modes have $5$ independent degrees of freedom, while the massless one has only $2$. For the latter, the $1/3$ factor in (\ref{Em}) must be replaced 
by $1/2$. Using Eqs.~(\ref{lambdawarp}, \ref{lambdaflat}) for the coupling $\lambda_{\bar{0}}$ of the massless graviton mode $m_{\bar{n}} = 0$, this gives
\be
\frac{d E_{\bar{0}}}{d\Omega_{\mathbf{k}}} = \frac{G}{2 \pi^2}\,\int dk\,k^2\,
\left(T^{\mu\nu}(k, \mathbf{k})\,T_{\mu\nu}^{*}(k, \mathbf{k}) - \frac{1}{2}\,|T^{\lambda}_{\lambda}(k, \mathbf{k})|^2\right)
\ee
for the energy radiated in the zero-mode per element of solid angle $d\Omega_{\mathbf{k}}$. This is the standard expression for the energy emitted in gravitational waves by a source, see e.g.~\cite{weinberg}~\footnote{Note that our convention (\ref{FT}) for the Fourier transform of the energy-momentum tensor with respect to space and time differs from the one used in \cite{weinberg} by a factor of $2\pi$.}. In the following, we will be interested in the production of massive modes 
$m_{\bar{n}} \neq 0$.

The perturbative production of massive scalar particles by cosmic string loops has been studied by several 
authors~\cite{sred, DVdil, PS, vach, sabancilar}. The calculation in our case is similar, except that we deal with massive spin-$2$ fields and that we have a tower of them. As in the case of moduli production~\cite{DVdil,PS}, KK modes of mass 
$m_{\bar{n}}$ are produced by loops of invariant length $L$ when the frequency $f \sim j / L$ associated to harmonics $j$ of the loop oscillation is greater than $m_{\bar{n}}$. This occurs for cusps and kinks with sufficiently high harmonics 
$j \gg 1$ throughout the cosmological evolution. This may also occur for low harmonics $j \sim 1$ for very small loops with 
$L < 1/m_{\bar{n}}$. Note however that such loops would be smaller than the effective size of the extra-dimensions. These loops may be relevant in the very early universe, just after the cosmic super-strings have been produced, although the motion of the strings may still be heavily damped at that epoch. In any case, during most of the cosmological evolution, the vast majority of loops satisfies $L \gg 1/m_{\bar{n}}$. We therefore focus on that case in the following.

To proceed, we need to calculate the Fourier transform (\ref{FT}) of the loop energy-momentum tensor (\ref{Tmunu})
\be
\label{TmunuFT}
T^{\mu\nu}(\omega_k, \mathbf{k}) = \mu\,\int d\tau d\sigma\,\sqrt{-\gamma}\,\gamma^{\alpha \beta}\,
\partial_\alpha X^\mu\,\partial_\beta X^\nu\,e^{i\,k_\lambda\,X^\lambda}
\ee
where $x^\lambda = X^\lambda(\tau, \sigma)$ denotes the four-dimensional coordinates of the loop. We follow the approach used in \cite{DV} for the production of gravitational waves. The main difference for the production of massive particles is that the four-vector $k^\lambda = (\omega_k, \mathbf{k})$ is now timelike instead of null. In addition, the massive 
spin-$2$ fields have $5$ degrees of freedom instead of $2$, as mentioned above. In the conformally flat gauge 
(\ref{gaugecons}), the energy-momentum tensor (\ref{TmunuFT}) can be written as~\cite{DV}~\footnote{In \cite{DV}, the time variation of the source is developed into a Fourier series, which leads to an extra factor of $1 / T = 2 / L$ in 
Eqs.~(\ref{FT}, \ref{TmunuDV}). Here we work instead with a continuous Fourier transform because this makes the notations somewhat simpler.}
\be
\label{TmunuDV}
T^{\mu\nu}(\omega_k, \mathbf{k}) = \frac{\mu}{2}\,I_+^{( \mu}\,I_-^{\nu )} \hspace*{0.5cm} \mbox{with} 
\hspace*{0.5cm} I_{\pm}^\mu = \int_{-\frac{L}{2}}^{\frac{L}{2}} d\sigma_\pm\,\dot{X}^\mu_\pm\,e^{\frac{i}{2}\,
k_\lambda X_\pm^\lambda}
\ee
where $I_+^{( \mu}\,I_-^{\nu )} = \left(I_+^\mu\,I_-^\nu + I_+^\nu\,I_-^\mu\right)/2$. The functions 
$X^{\mu}_{+}(\sigma_+)$ and $X^{\mu}_{-}(\sigma_-)$ of the worldsheet coordinates $\sigma_{\pm}$ are defined in 
Eq.~(\ref{Xpm}), and a dot on $X_+^\mu(\sigma_+)$ or $X_-^\mu(\sigma_-)$ denotes the derivative with respect to the corresponding unique variable, $\sigma_+$ or $\sigma_-$. Choosing a time gauge where 
$X^0_{\pm}(\sigma_\pm) = \sigma_{\pm}$, the dynamics of the loop is fully described by the two $3$-vectors $\mathbf{X}_+$ and $\mathbf{X}_-$. These are periodic functions of $\sigma_\pm$ of period $L$, where $L$ is the invariant length of the loop, and their derivative has unit norm: $|\dot{\mathbf{X}}_+| = |\dot{\mathbf{X}}_-| = 1$. 

For $m_{\bar{n}}\,L \gg 1$, the integrals $I_\pm^\mu$ can be calculated in the stationary phase approximation. The phase in $I_+^\mu$ is stationary when
\be
\label{stationary}
k_\lambda \, \dot{X}_+^\lambda = \sqrt{k^2 + m_{\bar{n}}^2} - k\,\cos\theta = 0
\ee
where we have used (\ref{4k}) and $\dot{X}_+^0 = |\dot{\mathbf{X}}_+| = 1$. Here and in the following, 
$\theta$ denotes the angle between $\mathbf{k}$ and $\dot{\mathbf{X}}_+$. We see from (\ref{stationary}) that the phase of 
$I_+^\mu$ is exactly stationary only for $m_{\bar{n}} = 0$ and $\theta = 0$. Nevertheless, the phase remains smaller than unity for some range of values of $m_{\bar{n}} \ll k \simeq \omega_k$ and $\theta \ll 1$ that we will determine below. Similarly, for the phase of $I_-^\mu$ to be stationary, $\mathbf{k}$ must be parallel to $\dot{\mathbf{X}}_-$. The vectors 
$\dot{\mathbf{X}}_+$ and $\dot{\mathbf{X}}_-$ are then parallel (and thus equal since they both have unit norm), 
which corresponds to a cusp. Another possibility is to have $\mathbf{k}$ parallel to $\dot{\mathbf{X}}_+$ and a discontinuity in $\dot{X}^\mu_-$, which corresponds to a kink. The effect of a kink is smaller than the one of a cusp, so we focus on cusps in the following~\footnote{We note however that kinks may play an important role too for loops with junctions where kinks proliferate~\cite{kinkprolif}.}. 

The phases in $I_\pm^\mu$ in the vicinity of a cusp, chosen to occur at $\sigma_\pm = 0$ and $X^\mu = 0$, are calculated in the Appendix. As in \cite{DV}, we take
\be
\label{ddotXL}
|\ddot{X}_\pm| \approx \frac{2\pi}{L} 
\ee
which is expected to be a good approximation for loops that are not too wiggly. We then obtain
\be
\label{phases}
\frac{1}{2}\,k_\lambda \,X_\pm^\lambda \, \approx \, 
\frac{k}{4}\,\,\left(\theta^2 + \frac{m_{\bar{n}}^2}{k^2}\right)\,\sigma_\pm + 
\frac{\pi^2\,k}{3\,L^2}\,\sigma_\pm^3 
\ee
where $m_{\bar{n}} \ll k \simeq \omega_k$ and $\theta \ll 1$. 

For KK modes emitted in the direction of the cusp velocity, $\theta = 0$, the phases (\ref{phases}) remain smaller than unity for $|\sigma_\pm| \lesssim 4\,k / m_{\bar{n}}^2$ and $|\sigma_\pm| \lesssim L / (\pi k L)^{1/3}$. The second bound is the most restrictive one for $k \gtrsim k_c$, where we define
\be
\label{kc}
k_c = \frac{1}{4}\,m_{\bar{n}}\,\sqrt{m_{\bar{n}}\,L} \, .
\ee
Thus for $k \gtrsim k_c$, the integrals $I_\pm^\mu$ are dominated by the range of worldsheet coordinates 
$|\sigma_{\pm}| \lesssim \Delta \sigma$, with
\be
\label{DeltaSigma}
\Delta \sigma \approx \frac{L}{(\pi k L)^{1/3}} \, .
\ee
In that case, the phases (\ref{phases}) remain also smaller than unity for $\theta \lesssim \theta_c$, where 
we define
\be
\label{thetac}
\theta_c = \frac{4^{2/3}}{(k L)^{1/3}}
\ee
and the numerical factor $4^{2/3}$ has been chosen for later convenience. For $\theta \gg \theta_c$ or $k \ll k_c$, the integrals $I_\pm^\mu$ are exponentially suppressed because the integrand oscillates rapidly. 

The full $\mathbf{k}$-dependence of the integrand in Eq.~(\ref{Em}) can be calculated analytically in the stationary phase approximation, and the integral over $d^3 \mathbf{k}$ can then be performed numerically. This is done in the Appendix, where we find
\be
\label{En}
E_{\bar{n}} \, = \, C \, \lambda_{\bar{n}}^2 \, \mu \, \sqrt{\frac{L}{m_{\bar{n}}}} 
\ee
for the energy emitted in KK modes with mode numbers $\bar{n}$, with $C \approx 0.4$ (the precise value depends on the shape of the cusp). The integral over $d^3 \mathbf{k}$ in (\ref{Em}) is dominated by the contribution from $k \sim k_c$ and $\theta \sim \theta_c$. Note that most of the KK modes are produced with a very large boost in the rest frame of the loop, 
$k / m_{\bar{n}} \, \sim \, \sqrt{m_{\bar{n}}\,L} / 4 \, \gg \, 1$. The number $N_{\bar{n}}$ of KK modes with mode numbers 
$\bar{n}$ produced by the cusp is obtained similarly by multiplying the integrand in (\ref{Em}) by $1 / \omega_k$. This gives
\be
\label{Nn}
N_{\bar{n}} \, = \, D \, \lambda_{\bar{n}}^2 \, \frac{\mu}{m^2_{\bar{n}}}
\ee
with $D \approx 0.3$.

We now discuss the range of parameters for which the calculation above is valid. It was shown in \cite{BPO} that, for cosmic strings of width $r$, the string segments on both sides of the cusp overlap in the region 
$\sigma \lesssim \sqrt{r\,L}$ around the cusp. For cosmic super-strings, this means that the distance between the two string segments in the rest frame of the cusp is shorter than the (local) string scale for 
$\sigma \lesssim \sqrt{L / M_s}$. The effective description of the loop dynamics based on the Nambu-Goto action would break down in this regime, so we must require the range $\Delta \sigma$ of values of $\sigma_\pm$ relevant for the above calculation to satisfy $\Delta \sigma > \sqrt{L / M_s}$. For $k > k_c$ and using Eq.~(\ref{DeltaSigma}), this gives
\be
\label{kMs}
k_c \, < \, k \, < \, \frac{M_s}{\pi}\,\sqrt{M_s\,L} \, . 
\ee
The constraint $k < M_s\,\sqrt{M_s\,L}$ also ensures that the energy of the KK modes is smaller than $M_s$ in the rest frame of the cusp. As shown in the Appendix, the calculation of the energy (\ref{En}) and number (\ref{Nn}) of KK modes with mode numbers $\bar{n}$ emitted by a cusp is dominated by a range of $\sigma$ values which is larger than 
(\ref{DeltaSigma}) evaluated at $k = k_c$. This is larger than $\sqrt{L / M_s}$ for $m_{\bar{n}} \lesssim M_s$. Thus the main constraint is that the calculation be limited to KK modes that are lighter than the fundamental (local) string scale $M_s$, which is consistent with neglecting fundamental string excitations~\footnote{The region around the cusp where the distance between the two string segments is shorter than the string scale may lead to the production of string states, which would then decay into KK modes. One can then expect this extra energy released in string states to be comparable to the one emitted by cusp annihilation on field theory cosmic strings~\cite{BPO}. As we will see, this is of the same order of magnitude as the total energy emitted directly into KK modes within the regime of validity of the Nambu-Goto description that we consider here. Therefore, the production of string states would not modify the order of magnitude of our results. At a more accurate level, however, one or the other effect may dominate, depending on the parameters.}. 

Up to now, we have considered the production of KK modes with given mode numbers $\bar{n}$ by a cusp. We now calculate the total energy emitted by the cusp in the form of massive KK modes. To do this, we have to sum Eq.~(\ref{En}) over all the KK modes of mass $m_{\bar{n}} < M_s$. This gives
\be
\label{Etot}
E \, = \, \sum_{\bar{n}}^{m_{\bar{n}} < M_s} E_{\bar{n}} \, = \, 
\frac{\kappa_E}{2}\,g_s^{2 - 5 \chi/4}\,\mu^{3/4}\,\sqrt{L}
\ee
where we have defined
\be
\label{kappaE}
\kappa_E \, \equiv \, \frac{2\,C}{(2 \pi)^{1/4}} \, 
\left(\frac{\sqrt{\alpha'}}{R}\right)^{d-1/2}\,\sum_{\bar{n}}^{m_{\bar{n}} < M_s} c_{\bar{n}}^{-1/2}
\ee
and we have used Eqs.(\ref{muwarp}, \ref{muflat}), (\ref{lambdawarp}, \ref{lambdaflat}) and (\ref{moverM}). 
The coeffficient $\kappa_E$ depends on the details of the compactification and of the KK spectrum. As discussed below 
Eq.~(\ref{moverM}), Eqs.~(\ref{Etot}, \ref{kappaE}) apply both to the case of a flat internal space with $d$ large extra dimensions (of radius $R \gg \sqrt{\alpha'}$) and, for $d = 6$, to the case of a warped throat (with a radius of curvature 
$\tilde{R} = 2\pi\,R$ at the bottom of the throat). 

In the case of toroidal compactification with $d$ large extra dimensions discussed in sub-section~\ref{flat}, we have 
$c_{\bar{n}} = |\bar{n}| = \sqrt{n_1^2 + ... + n_d^2}$, see Eq.~(\ref{mflat}). In that case, many KK modes are lighter than the string scale and we calculate the sum over $\bar{n}$ in Eq.~(\ref{kappaE}) by approximating it with an integral 
\be
\label{sumn}
\sum_{\bar{n}}^{m_{\bar{n}} < M_s} c_{\bar{n}}^{-1/2} \, \approx \, \int_{|\bar{n}| < \frac{R}{\sqrt{\alpha'}}} d^d\bar{n}
\,|\bar{n}|^{-1/2} \, = \, \frac{S_{d-1}}{d-1/2}\,\left(\frac{R}{\sqrt{\alpha'}}\right)^{d-1/2} 
\ee
where $S_{d-1}$ is the area of the $(d-1)$-dimensional unit sphere. Eq.~(\ref{kappaE}) then shows that $\kappa_E$ is independent of $R / \sqrt{\alpha'}$, with $\kappa_E \approx 2 - 3$ for $d = 1 - 6$. Indeed, when $R / \sqrt{\alpha'}$ increases, the coupling $\lambda_{\bar{n}}$ of the cosmic string to a given KK mode decreases, but the number of KK modes whose mass is smaller than $M_s$ increases. Correspondingly, the factor in front of the the sum in (\ref{kappaE}) decreases, but more terms must be included in the sum. These two effects compensate each other.  

In the case of a warped throat, we must use Eq.~(\ref{kappaE}) with $d = 6$. The detailed KK spectroscopy for a compact throat is in general not known. However, for a Klebanov-Strassler throat approximated as a patch of $AdS_5 \times T^{1,1}$, 
we can estimate $\kappa_E$ by using Table 1 of \cite{KKrelics} for the spectrum of the lowest-lying spin-$2$ KK modes. We find again that $\kappa_E$ does not significantly depend on $\tilde{R} / \sqrt{\alpha'}$, with $\kappa_E \approx 10$ for a range of values around $\tilde{R} / \sqrt{\alpha'} \sim 10$. This is of the same order of magnitude as the value that we obtained above for a flat internal space with large volume. Note that, in the case of a warped throat, the characteristic scale $\tilde{R} = 2 \pi\,R$ for the curvature of the throat (\ref{dsapprox}) should be larger than $\sqrt{\alpha'}$ in order to trust the super-gravity approximation, but it is not expected to be much larger. Thus a relatively small amount of KK modes are lighter than the string scale, but their coupling to the cosmic strings is much stronger than in the case of a flat internal space with large volume. We see that $\kappa_E$ and the the total energy (\ref{Etot}) emitted in massive KK modes by a cusp are comparable in both cases. 

The total number of KK modes emitted by a cusp can be calculated in a similar way, by summing (\ref{Nn}) over all the modes with $m_{\bar{n}} < M_s$. This gives
\be
\label{Ntot}
N \, = \, \sum_{\bar{n}}^{m_{\bar{n}} < M_s} N_{\bar{n}} \, = \, \kappa_N \, g_s^{2 - 2 \chi} 
\ee
where we have defined
\be
\kappa_N \, = \, \frac{D}{2 \pi} \, 
\left(\frac{\sqrt{\alpha'}}{R}\right)^{d - 2} \, \sum_{\bar{n}}^{m_{\bar{n}} < M_s} \, c_{\bar{n}}^{-2} 
\ee
and we have used Eqs.~(\ref{muwarp}, \ref{muflat}), (\ref{lambdawarp}, \ref{lambdaflat}) and (\ref{moverM}). 

In the case of toroidal compactification with $d$ large extra dimensions, we can again approximate the sum by an integral as in (\ref{sumn}). For $d \geq 3$, we find again that $\kappa_N$ is independent of $R / \sqrt{\alpha'}$, with 
$\kappa_N \approx 0.4 - 0.6$ for $d = 3 - 6$. However, for $d = 2$, $\kappa_N$ now depends logarithmically on 
$R / \sqrt{\alpha'}$. Furthermore, in the case of only one large extra dimension, $d = 1$, the sum over the (in that case single) quantum number $n$ is dominated by the minimal value $n = 1$, and $\kappa_N$ is proportional to 
$R / \sqrt{\alpha'}$. Finally, in the case of a Klebanov-Strassler throat, we estimate $\kappa_N$ by using the results of \cite{KKrelics} as we did for $\kappa_E$. We find $\kappa_N \sim 3$ for a range of values around 
$\tilde{R} / \sqrt{\alpha'} \sim 10$. Putting all this together, the total number of KK modes emitted by the cusp is 
given by
\be
\label{Ntot2}
N \, \approx
\left\{
\begin{array}{l}
0.1 \, g_s^{2 - 2 \chi}\,\frac{R}{\sqrt{\alpha'}} 
\hspace*{1.5cm} \mbox{ for } d = 1\\
0.1 \, g_s^{2 - 2 \chi}\,\mathrm{ln}\left(\frac{R}{\sqrt{\alpha'}}\right) 
\hspace*{0.5cm} \mbox{ for } d = 2\\
\kappa_N \, g_s^{2 - 2 \chi} 
\hspace*{2.2cm} \mbox{ for } d \geq 3 \mbox{ and warped throat.}
\end{array}
\right.
\ee
Note that, for $d = 1$ large extra-dimension compactified on a torus, $R \gg \sqrt{\alpha'}$ and $N$ is strongly enhanced. In that case, many light KK modes with $m_{\bar{n}} \sim 1/R$ are emitted by the cusp. However, most of the energy is still carried away by a small amount of much heavier KK modes with $m_{\bar{n}} \sim M_s$, so that the total energy (\ref{Etot}) that is emitted is of the same order of magnitude as in the cases with $d \geq 2$.  

The power emitted in massive KK modes by a loop of length $L$ can be calculated as $P_{KK} \approx c\,E / T$, where 
$c$ is the average number of cusps on loops per oscillation period (expected to be $c \approx 1$) and $T = L / 2$ is the period of oscillation. Using (\ref{Etot}), this gives
\be
\label{Pkk}
P_{KK} = \Gamma_{KK}\,\frac{\mu^{3/4}}{\sqrt{L}}
\ee
with 
\be
\label{Gammakk}
\Gamma_{KK} \, \approx \, c \, \kappa_E \, g_s^{2 - 5 \chi / 4} \, .
\ee
For $c \approx 1$, $\kappa_E \approx 10$, and $g_s \approx 0.1 - 1$, this gives $\Gamma_{KK} \approx 0.1 - 10$. 

Interestingly, the power emitted in KK modes by cosmic super-strings is comparable the one emitted by standard Abelian-Higgs cosmic strings in their constituent field through the process of cusp annihilation. When two segments of a field theory cosmic string overlap, their stability is not anymore protected by topology and they can annihilate into radiation of the fields that constitute the cosmic string. This occurs in particular around a cusp, where string segments of invariant width $r$ overlap in the region $\sigma_c \sim \sqrt{r\,L}$ and release the corresponding energy $\mu\,\sigma_c$ in the form of (highly boosted) particles of the constituent fields~\cite{BPO}. For $r \sim \mu^{-1/2}$ and $c \sim 1$ cusp per oscillation period, the power emitted in massive particles is~\footnote{Abelian-Higgs cosmic strings of the extreme type-I kind have a larger width $r$ so that $\Gamma_{part} \gg 1$ in that case, see e.g.~\cite{flatdir}.}
\be
\label{Ppart}
P_{\mathrm{cusp\,annihil}} \, \sim \, P_{KK} \, \sim \, \frac{\Gamma_{part}\,\mu^{3/4}}{\sqrt{L}} \equiv P_{part}
\ee
where $\Gamma_{part} \sim 1$. Similarly, cusp annihilation emits about $1$ particle of the constituent fields, which is comparable to (\ref{Ntot2}) for $d > 2$, up to factors of the string coupling $g_s$. Indeed, whether for KK emission by cosmic super-strings or for cusp annihilation on field theory cosmic strings, the dynamics of the cusp is the same in both cases and there are only two independent scales in the problem ($\mu$ and $L$), so it is not surprising a posteriori that the results are similar in both cases. Note however that we obtained our results for KK emission by a perturbative calculation within the regime of validity of the effective Nambu-Goto description. By contrast, cusp annihilation is a non-perturbative process that occurs beyond the Nambu-Goto approximation. We will also see in the following that KK emission by cosmic super-strings may have specific cosmological consequences.

\section{Consequences for the Loop Number Density and the Friction-Dominated Epoch}
\label{distribution}

Before addressing cosmological consequences of KK emission, we must study how it modifies the number density of loops in the universe. Since we saw above that the power emitted in KK modes by cosmic super-strings is comparable to the one emitted by cusp annihilation on standard Abelian-Higgs cosmic strings, the effects on the loop number density are the same in both cases. In this Section, we will therefore call "massive particles", or simply "particles", the quanta produced by cusps, whether these are KK modes or quanta of the constituent fields of Abelian-Higgs cosmic strings.

In addition to massive particles produced at cusps, cosmic string loops emit gravitational waves throughout their oscillations, with the power
\be
\label{Pgrav}
P_{grav} = \Gamma\,G \mu^2
\ee
where $\Gamma \sim 50$ depends on the loop oscillation (see e.g.~\cite{Vachaspati:1984gt}). Comparing (\ref{Pgrav}) with (\ref{Ppart}), we see that loops of size
\be
\label{L=}
L_= = \frac{\Gamma^2_{part}}{(\Gamma G \mu)^2\,\sqrt{\mu}}
\ee
emit the same amount of energy in massive particles and in gravitational radiation. Particle emission dominates for smaller loops. Compared to the case where only gravitational radiation is taken into account, the loops with $L < L_=$ will thus have a shorter lifetime, and therefore also a smaller number density. Such loops are relevant mainly at early times or for small string tensions $\mu$.

After a possible early friction-dominated evolution (where the motion of the strings is heavily damped due to their interactions with particles in the high-density thermal plasma~\cite{friction1, friction2}), a network of cosmic strings enter a scaling regime where its energy density remains a fixed fraction of the total energy density of the universe. The effect of the friction-dominated epoch, in particular for cosmic super-strings, will be discussed in sub-section~\ref{friction}. In the scaling regime, the network of long strings loses energy by producing loops at the rate~\cite{VS}
\be
\label{drholoop}
\frac{d \rho_{\mathrm{loop}}}{d t} \approx \frac{\zeta\,p^{-\beta}\,\mu}{t^3}
\ee 
when loop production is indeed responsible for maintaining the scaling regime. Numerical simulations give $\zeta \sim 10$ in the radiation-dominated epoch~\footnote{The value of $\zeta$ can be an order of magnitude smaller for loops produced during the matter-dominated era. In the following, we will only be interested in loops produced during the radiation-dominated era.}. For cosmic super-strings, the reconnection probability $p$ can be smaller than unity. This is expected to increase the number density of strings, and we took this effect into account in (\ref{drholoop}) through the factor 
$p^{-\beta}$. The exponent $\beta$ is usually expected to be $\beta = 1$~\cite{DVsuper}, but the values 
$\beta = 2$~\cite{tye} and $0.6$~\cite{avgou} have also been obtained. We therefore keep $\beta$ as a parameter here. Typical values of the reconnection probabilities obtained in \cite{psuper} are in the ranges 
$10^{-3} \lesssim p \lesssim 1$ for F-strings and $10^{-1} \lesssim p \lesssim 1$ for D-strings.

The loop number density depends crucially on the scale of loop production from the network of long strings. This issue is still under debate, see e.g.~\cite{alpha1, alpha2, alpha3, simul1, simul2}. According to the most recent simulations~\cite{simul1, simul2}, the characteristic initial size $L_i$ of the loops depends on the cosmic time $t$ when they are produced as
\be
\label{Li}
L_i = \alpha\,t
\ee
with $\alpha \sim 0.1$. This is the picture that we will adopt in the following. In that case, the initial size of the loops depends on the network dynamics on large scales and not on the energy radiated by the long strings on smaller scales. Thus, even if the long strings themselves emit massive particles too, this should not modify the initial size of the loops.

\subsection{Loop Number Density in the Scaling Regime}

Consider first the rate at which a loop shrinks once it is produced. Its instantaneous energy $E = \mu L$ is radiated in gravitational waves and in massive particles at the rates (\ref{Pgrav}) and (\ref{Ppart}), respectively. Its length then varies as
\be
\label{Ldot}
\dot{L} = - \Gamma\,G \mu\,\left(1 + \sqrt{\frac{L_=}{L}}\right)
\ee 
where a dot denotes derivative with respect to cosmic time $t$ and we have used Eq.~(\ref{L=}). The lifetime of a loop of length $L > L_=$ is dominated by the time it spends shrinking from $L$ to $L_=$ by emitting mostly gravitational waves:
$\Delta t \sim L / (\Gamma G \mu)$. This is larger than the Hubble time $t$ for $L > \Gamma G \mu\,t$. On the other hand, 
a loop of length $L < L_=$ shrinks to zero size by emitting mostly massive particles, in a time 
$\Delta t \sim \mu^{1/4}\,L^{3/2} / \Gamma_{part}$. This is larger than the Hubble time for $L > L_{part}(t)$, where 
\be
\label{Lpart}
L_{part}(t) = \frac{(\Gamma_{part}\,t)^{2/3}}{\mu^{1/6}} \, .
\ee
Thus in both cases, loops survive during more than a Hubble time if their length statisfy $L > L_H(t)$, where
\be
\label{LHt}
L_H(t) = \mathrm{Max}\left[\Gamma G \mu t , L_{part}(t)\right]
\ee
with $\mathrm{Max}[a , b] = a$ or $b$, whichever is larger. 

There is a unique time $t_{\mu}$ where the three length scales $L_=$, $L_{part}(t)$ and $\Gamma G \mu\,t$ are all equal to each other
\be
\label{tmu}
t_\mu \, = \, \frac{\Gamma^2_{part}}{(\Gamma\,G \mu)^3\,\sqrt{\mu}} 
\, = \, \frac{\Gamma^2_{part}}{\Gamma^3}\,\frac{\tp}{(G \mu)^{7/2}} 
\ee
where $\tp \simeq 5.4 \times 10^{-44}$ sec is the Planck time. Note that $L_= < L_{part} < \Gamma G \mu\,t$ for 
$t > t_{\mu}$, while $\Gamma G \mu\,t < L_{part} < L_=$ for $t < t_{\mu}$. For $t > t_{\mu}$, the loops are long-lived compared to the Hubble time if they are produced with an initial length $L_i = \alpha t > L_H(t) = \Gamma G \mu\,t$, which is always satisfied for $\alpha \sim 0.1$. For $t < t_{\mu}$, the loops are long-lived if $L_i = \alpha\,t > L_{part}(t)$, which gives $\alpha > \Gamma_{part} / \sqrt{\mu^{1/2}\,L_i}$. Again, this is always satisfied for $\alpha \sim 0.1$, because $\Gamma_{part} \sim 1$ and $\mu^{1/2}\,L_i \gg 1$. Thus in either case, the loops survive for more than one Hubble times after their production. At a given time $t$, there is a distribution of loops with all lengths $L \leq \alpha\,t$. Of particular interest are the loops with length $L = L_H(t)$, which are going to die in about one Hubble time. For 
$t > t_{\mu}$, integrating Eq.~(\ref{Ldot}) between $L_H(t) = \Gamma G \mu t$ and $L_i = \alpha t_i$ gives 
$L_i - L_H(t) \sim L_H(t)$ for $t \gg t_i$. For $t < t_{\mu}$, integrating Eq.~(\ref{Ldot}) between $L_H(t) = L_{part}(t)$ and $L_i = \alpha t_i$ gives $L^{3/2}_i - L^{3/2}_H(t) \sim L^{3/2}_H(t)$. This shows that, in both cases, the loops of length $L = L_H(t)$ at time $t$ where produced at the time $t_i \sim L_H(t) / \alpha \ll t$.

The quantity of interest is the number density of loops of length $L$ at time $t$, $n_L(t)$. This is defined so that the number of loops with lengths between $L$ and $L + dL$ per unit physical volume at time $t$ is $n_L(t)\,dL$. A detailed study of $n_L(t)$ at all lengths $L$ for loops emitting both gravitational waves and massive particles will appear 
elsewhere. Here we will only present the results that we will need in the next Section. During the radiation-dominated era ($t < t_{eq} \simeq 1.8 \times 10^{12}$ sec), the loop number density can be written as
\be
\label{nrad}
\mbox{For } t < t_{eq}: \hspace*{0.5cm} n_L(t) \, \approx \, 
\frac{\zeta\,p^{-\beta}\,\sqrt{\alpha}}{L^{5/2}\,t^{3/2}} \hspace*{0.5cm} \mbox{ for } \hspace*{0.2cm} 
L_H(t) < L < \alpha\,t 
\ee
where we assume that all the loops are produced from the network of long strings with the same initial size 
$L_i = \alpha\,t$ at time $t$. The loops with length $L = L_H(t)$ have a lifetime of the order of the Hubble time and are the most abundant ones. The loops with length $L < L_H(t)$ have a number density that is further reduced by particle emission and they will not be relevant for us. For $t > t_{\mu}$, Eq.~(\ref{LHt}) gives $L_H(t) = \Gamma G \mu t$ and 
Eq.~(\ref{nrad}) reduces to the standard result for loops emitting only gravitational waves, see e.g.~\cite{VS}. On the other hand, for times and string tensions such that $t < t_{\mu}$, most of the loops decay by emitting mainly particles, so their lifetime and their number density is reduced. 

Consider now the loop number density during the matter-dominated era, at a time $t > t_{eq}$. If loops produced during the 
radiation-dominated era survive until the time $t$, they dominate the loop number density at that time. This occurs if 
$L_H(t) < \alpha t_{eq}$. For $t < t_{\mu}$, this condition reads $L_{part}(t) < \alpha t_{eq}$. For 
$\Gamma_{part} \sim 1$ and $\alpha \sim 0.1$, this is satisfied until today ($t = t_0 \simeq 4.3 \times 10^{17}$ sec) when 
$G \mu \gtrsim 10^{-93}$. Thus, when $t_0 < t_\mu$, loops produced during the radiation dominated era survive until today for any realistic value of the string tension. On the other hand, for $t > t_{\mu}$ the condition $L_H(t) < \alpha t_{eq}$ reads $\Gamma G \mu t < \alpha t_{eq}$. For $\Gamma \sim 50$ and $\alpha \sim 0.1$, this is satisfied until today for 
$G \mu \lsim 10^{-8}$, which will hold for the cosmological applications we will consider in the next Section. Thus in both cases, the loop number density during the matter era is dominated by loops that have been produced during the radiation era. It can be written as
\be 
\label{nmat}
\mbox{For } t > t_{eq} \hspace*{0.1cm} \mbox{ and } \hspace*{0.1cm} L_H(t) < \alpha t_{eq} :
\hspace*{0.5cm} n_L(t) \, \approx  \, \frac{\zeta\,p^{-\beta}\,\sqrt{\alpha\,t_{eq}}}{L^{5/2}\,t^2} \hspace*{0.5cm} 
\mbox{ for } \hspace*{0.2cm} L_H(t) < L < \alpha\,t_{eq} 
\ee
where the loops with length $L = L_H(t)$ are again the most abundant ones. As was the case in the radiation era, 
Eq.~(\ref{nmat}) reduces to the standard result for loops emitting only gravitational waves when $t > t_{\mu}$, while the loop number density is further suppressed by particle emission when $t < t_{\mu}$.  

In the following Section, we will be interested in the total energy density in KK modes that is emitted by the population of loops per unit time. This can be calculated as
\be
\label{dotrhokk}
\dot{\rho}_{KK}(t) \, = \, \int dL\,n_L(t)\,P_{KK}
\ee
where $P_{KK}$ is the power (\ref{Pkk}) emitted in KK modes by a loop of length $L$, and we now set 
$\Gamma_{part} = \Gamma_{KK}$. The integral over loop lengths is dominated by the contribution of $L \approx L_H(t)$. In the radiation-dominated era, this gives
\be
\label{drhorad}
\dot{\rho}_{KK}(t) \, \approx \, \frac{\Gamma_{KK}\,\mu^{3/4}\,\zeta\,p^{-\beta}\,\sqrt{\alpha}}{(\Gamma G \mu)^2\,t^{7/2}}\,\,\mathrm{Min}\left[1 \, , \, \frac{t^{2/3}}{t_{\mu}^{2/3}}\right] \hspace*{0.8cm} \mbox{ for } t < t_{eq}
\ee
where we have used (\ref{nrad}) with (\ref{LHt}) and (\ref{tmu}). In the matter-dominated era, we have instead
\be
\label{drhomat}
\dot{\rho}_{KK}(t) \, \approx \, \frac{\Gamma_{KK}\,\mu^{3/4}\,\zeta\,p^{-\beta}\,\sqrt{\alpha\,t_{eq}}}{(\Gamma G \mu)^2\,t^4}\,\,\mathrm{Min}\left[1 \, , \, \frac{t^{2/3}}{t_{\mu}^{2/3}}\right] \hspace*{0.8cm} \mbox{ for } t > t_{eq}
\ee
where we have used (\ref{nmat}). In these two equations, $\mathrm{Min}[a , b] = a$ if $a < b$ and $b$ if $b < a$. 

For the cosmological consequences to be discussed in the next Section, it is important to note that 
$\dot{\rho}_{KK} \propto \mu^{-5/4}$ for $t > t_{\mu}$ while $\dot{\rho}_{KK} \propto \mu^{13/12}$ for $t < t_{\mu}$. At a given time $t$, if the string tension is sufficiently large so that $t > t_{\mu}$, most of the loops decay by emitting mainly gravitational waves. Decreasing $\mu$ then increases the loop lifetime, and thus also the loop number density and 
$\dot{\rho}_{KK}$. However, when $\mu$ becomes sufficicently small, $t_{\mu}$ becomes smaller than $t$. For $t < t_{\mu}$, the decay of the loops is dominated by KK emission. Decreasing $\mu$ then \emph{decreases} the lifetime of the loops, and thus also their number density and $\dot{\rho}_{KK}$.

\subsection{Consequences of the Friction-Dominated Epoch}
\label{friction}

We now discuss the effect of the friction-dominated epoch of cosmic string evolution on the loop distribution considered in the previous sub-section.  

After their formation at time $t_f \sim \tp / (G \mu)$, the motion of field theory cosmic strings is usually heavily damped due to their interactions with the high-density thermal plasma of the universe~\cite{friction1, friction2}. This friction-dominated, or damped evolution ends at the time~\cite{VS}
\be
\label{tdFT}
t_d \sim \frac{\tp}{(G \mu)^2} \hspace*{0.5cm} \mbox{(field theory cosmic strings)}\, .
\ee
After that time, the strings start to acquire relativistic velocities. 

On the other hand, the damping of the motion of cosmic super-strings has not been studied so far. Cosmic super-strings are more likely to be stable if they are somehow decoupled from other degrees of freedom, in particular from the Standard Model fields~\cite{CMP}. Therefore, their interactions with the standard thermal plasma, and the resulting friction force, may be much less efficient. However, cosmic super-strings interact generically with KK modes, as we have seen. In particular, KK modes are abundantly produced by brane / anti-brane annihilation at the end of brane inflation, and they are then expected to quickly thermalize~\cite{warpreh, KKrelics}. Alternatively, KK modes may be produced thermally in the early universe at high temperatures, since we saw that their mass is smaller than the temperature $T \sim \sqrt{\mu}$ at which cosmic 
super-strings form. We then expect the motion of the strings to be damped by their interactions with this thermal gas of 
KK modes. This should occur at least in the case of a warped throat, where the KK modes are strongly coupled to the cosmic strings, as we discussed in sub-section~\ref{throat}. However, this damping can be efficient only for a relatively short amount of time. Indeed, when the temperature of the universe drops below the mass of the lowest-lying KK mode 
($\bar{n} = \bar{1}$ in (\ref{mwarp})), all the KK modes become non-relativistic. After that time, their number density is exponentially (Boltzmann) suppressed, so their interactions with the cosmic super-strings should become inefficient. The lowest-lying KK modes become non-relativistic at the time $t \sim \Mp / T^2$ when the temperature is 
$T \sim m_{\bar{1}} \sim e^{A_b} / \tilde{R}$. Using (\ref{muwarp}), this gives
\be
\label{tdST}
t_d \sim \left(\frac{\tilde{R}}{\sqrt{\alpha'}}\right)^2\,\frac{\tp}{G \mu} \hspace*{0.5cm} 
\mbox{(cosmic super-strings in a warped throat)}
\ee
for the end of the friction-dominated era in that case. This estimate is model-dependent and should be checked more carefully. However, comparing with (\ref{tdFT}) and remembering that $\tilde{R} / \sqrt{\alpha'}$ should not be parametrically large in a warped throat, we see that the friction-dominated era may be significantly shorter for cosmic super-strings than for field theory cosmic strings.

In the previous sub-section, we neglected the effect of friction. For short-lived loops, this is valid as soon as 
$t > t_d$. For long-lived loops, however, we must further require that the loops present at a given time $t$ were produced at a time $t_i$ after the friction-dominated epoch, $t_i > t_d$. For $t > t_{\mu}$, we saw that the loop number density is dominated by the loops with $L = \Gamma G \mu\,t$, which are produced at the time $t_i = \Gamma G \mu\,t / \alpha$. Using 
(\ref{tmu}) with $\alpha \sim 0.1$, $\Gamma_{part} \sim 1$ and $\Gamma \sim 50$, one finds that $t_i > t_d$ is always satisfied for $t > t_\mu$, $t_d \leq \tp / (G \mu)^2$ and $G \mu < 10^{-5}$. Therefore, the total number density of loops and $\dot{\rho}_{KK}$ are not affected by the friction-dominated epoch when $t > t_{\mu}$. This epoch can only affect the total number density in the regime $t < t_{\mu}$, where particle emission dominates over gravitational emission.

For $t < t_{\mu}$, we saw that the loop distribution at a given time $t$ is dominated by the loops with $L = L_{part}(t)$, which are produced at the time $t_i = L_{part}(t) / \alpha$. This is greater than $t_d$ at times $t > t_*$, where
\be
\label{tstar}
t_* \sim \frac{\alpha^{3/2}\,\mu^{1/4}\,t_d^{3/2}}{\Gamma_{part}} \, .
\ee
At times $t > t_*$, the friction-dominated epoch does not affect the total number density of loops and $\dot{\rho}_{KK}$. On the other hand, at times $t < t_*$, loops of size $L = L_{part}(t)$ are not present since they would have been produced before $t_d$. At these times, there are only larger loops, which have a smaller number density, so that $\dot{\rho}_{KK}$ would be reduced compared to the result given in Eqs.~(\ref{drhorad}, \ref{drhomat}).

\section{Constraints from Big-Bang Nucleosynthesis, the Diffuse Gamma-Ray Background and GW Experiments}
\label{constraints}

We now address cosmological consequences of KK emission by cosmic super-strings. 

Once produced by cusps, the fate of the KK modes depends \emph{a priori} on their couplings to the Standard Model fields and other degrees of freedom, which in turn depends on how the Standard Model (SM) is realized in the internal space. For instance, in the case of a warped throat with the SM fields localized on a brane, the coupling between the KK modes and the SM depends exponentially on the proper distance between the brane and the bottom of the throat (where the cosmic strings and the KK modes are localized). In general, however, the KK modes should decay mostly and relatively quickly into Standard Model (SM) degrees of freedom. Indeed, because the KK modes are light, they are also abundantly produced in the early universe at high temperatures. For instance, they are the main decay products of brane / anti-brane annihilation at the end of brane inflation, and their energy must then be efficiently converted into SM degrees of freedom in order to reheat the universe~\cite{warpreh, KKrelics}. These KK modes produced at early times must decay mostly into SM fields, and at the very last before the onset of Big-Bang Nucleosynthesis, for the scenario to be cosmologically viable. 

On the other hand, cusps on cosmic super-strings emit KK modes throughout the evolution of the universe, until the present epoch. The energy released in the cosmological medium when these KK modes decay into SM particles is then constrained by observations. More precisely, when the SM is realized on some brane in the internal space, the KK modes should decay mainly into photons and gluons, see~\cite{lykken} for the branching ratios of KK modes into the different SM species 
localized on a $3$-brane. The injection of such particles in relatively recent cosmological epochs is severely constrained by different cosmological probes, in particular Big-Bang Nucleosynthesis (BBN), the Cosmic Microwave Background (CMB) and the diffuse gamma-ray background. We now study how this constrains cosmic super-strings. Similar constraints have been studied for topological defect models of ultra-high energy cosmic rays in \cite{instant} and for Abelian-Higgs cosmic strings in the extreme type I regime in \cite{flatdir, flatdir2}. 

In the following, we will assume that, once produced by cusps, the KK modes decay in less than one Hubble time. This would certainly be true for KK modes produced at rest in the cosmological frame. Indeed, the lifetime at rest $\tau_r$ of most KK modes must be smaller than at least the time when BBN starts, $\tau_r < 1$ sec, while in the following we will be interested in later times, $t \gtrsim 10^8$ sec. In general, however, some of the KK modes can be long-lived, see e.g.~\cite{KKrelics}. Furthermore, the KK modes emitted by cusps are highly boosted in the cosmological frame, as discussed in Section~\ref{emission}. Their decay can then be delayed, until their energy has been sufficiently redshifted by the expansion of the universe. In that case, KK modes decaying at a given time have been produced earlier, when the loop number density was much greater, so that much more KK mode energy was produced. The energy density of the KK modes is then diluted by the expansion until they decay, but the net effect is still an increase of the energy density of the KK modes when they decay~\cite{letter}. This would then lead to stronger constraints compared to the ones we will obtain below by assuming that the KK modes decay in less than one Hubble time. However, at times $t \gtrsim 10^8$ sec, this occurs only for relatively large values of $\tau_r$, not much below $1$ sec, which is strongly model-dependent. In order to obtain conservative constraints, we will therefore not consider such a possibility.

The most stringent constraints on energy injection after recombination, at cosmic times $t \gsim 10^{13}$ sec, come from observations of the diffuse gamma-ray background, see e.g.~\cite{gammaray1, gammaray2}. Once the KK modes decay into photons and gluons, an electromagnetic cascade is very quickly established. The gluons hadronize and produce jets composed mainly of pions and a smaller fraction of nucleons. The neutral pions decay into photons and the charged ones into muons and neutrinos. The muons then further decay into electrons, positrons and neutrinos. The photons, electrons and positrons thus produced experience very fast cascade interactions with the cosmological photon background, until the energy of the injected photons falls below the threshold for electron-positron pair production. As a result, most of the energy injected into the cosmological medium by the decay of the KK modes is quickly re-processed into a diffuse flux of gamma-rays with energies below $100$ GeV~\cite{Berezinsky:1975zz}, which is constrained by the EGRET~\cite{egret} and Fermi-LAT~\cite{fermi} experiments.  

The cascade energy density accumulated up to the present epoch ($t = t_0 \simeq 4.3 \times 10^{17}$ sec) can be calculated 
as
\be
\label{wcas}
\omega_{cas} \, \approx \, f_{em}\,\int_{t_{cas}}^{t_0} dt\,\dot{\rho}_{KK}(t)\,\frac{a^4(t)}{a^4(t_0)}
\ee
where $f_{em} \sim 0.5$ is the fraction of energy density that goes into the electromagnetic cascade, and the last factor comes from the redshift of the photon energy density from the time of production until today. The lower limit in the time integral above comes from the fact that photons produced at too early times are absorbed by the cosmological medium. Following \cite{sabancilar}, we neglect the contribution of the epochs with redshifts $z > z_{cas} \sim 60$ to the cascade, which gives $t_{cas} \sim 10^{15}$ sec in Eq.~(\ref{wcas})~\footnote{The precise value of $t_{cas}$ is in any case not very important here, because it only affects the upper range of string tensions constrained by the diffuse gamma-ray background, where the constraints from BBN are stronger anyway. For the lower range of string tensions constrained by the diffuse gamma-ray background, the time integral in Eq.~(\ref{wcas}) is dominated by its upper limit.}. Neglecting the late-time acceleration of the universe expansion, we have $a(t) \propto t^{2/3}$ during the matter-dominated era and we can use Eq.~(\ref{drhomat}) for the rate of energy density emitted in KK modes, $\dot{\rho}_{KK}(t)$. The time integral in (\ref{wcas}) is then dominated by its lower limit for $t_{\mu} < t_{cas}$, by its upper limit for $t_{\mu} > t_0$, and by the contribution of $t = t_{\mu}$ for 
$t_{cas} < t_{\mu} < t_0$. This gives
\be
\omega_{cas} \, \approx \, 
\frac{\Gamma_{KK}\,\mu^{3/4}\,\zeta\,p^{-\beta}\,\sqrt{\alpha\,t_{eq}}}{(\Gamma G \mu)^2\,t_0^{8/3}\,t_{\mu}^{1/3}}\,
\,\mathrm{Min}\left[\frac{t_{\mu}^{1/3}}{t_{cas}^{1/3}} \, , \, 
2 - \frac{t_{cas}^{1/3}}{t_{\mu}^{1/3}} - \frac{t_{\mu}^{1/3}}{t_0^{1/3}} \, , \, 
\frac{t_0^{1/3}}{t_{\mu}^{1/3}}\right] \, .
\ee

\begin{figure}[htb]
\begin{center}
\includegraphics[width=10cm]{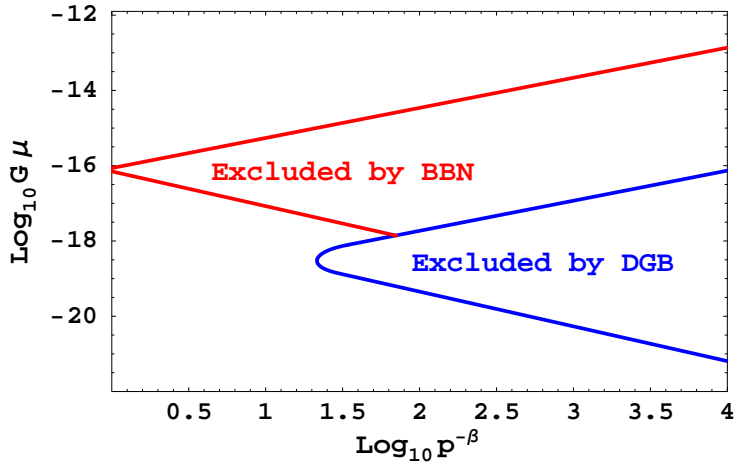}
\end{center}
\vspace*{-5mm}
\caption{Constraints on cosmic super-strings from KK emission in the $(p^{-\beta}, G \mu)$-plane, for $\zeta = 10$, 
$\alpha = 0.1$, $\Gamma = 50$ and $\Gamma_{KK} = 1$. The region inside the two straight lines (in red) in the upper-half of the plot is excluded by BBN (photo-dissociation of $^4 He$). The region inside the parabola-like curve (in blue) in the lower-half of the plot is excluded by observations of the diffuse gamma-ray background (DGB).}
\label{Ga1}
\end{figure}

Recent results from the Fermi-LAT experiment~\cite{fermi} lead to the constraint~\cite{wcas} 
$\omega_{cas} < \omega_{cas}^{max} \simeq 5.8 \times 10^{-7} \mathrm{eV} / \mathrm{cm}^3$. 
The resulting constraint on cosmic super-strings in the $(p^{-\beta}, G \mu)$-plane is shown in Fig.~\ref{Ga1} for 
$\zeta = 10$, $\alpha = 0.1$, $\Gamma = 50$ and $\Gamma_{KK} = 1$. The region inside the parabola-like curve (in blue) in the lower-half of the plot is excluded by observations of the diffuse gamma-ray background (DGB). As the reconnection probability $p$ decreases, the loop number density and the total energy density emitted in KK modes increase, so the constraint becomes stronger. For a given value of $p$, a range of tensions $\mu$ is excluded. In the upper-half of this range, the loop number density is dominated by gravitational emission, as discussed below Eq.~(\ref{drhomat}). In this regime, increasing $\mu$ decreases the loop number density and the total energy density emitted in KK modes, so that 
$\omega_{cas} < \omega_{cas}^{max}$ for sufficiently large tensions. On the other hand, in the lower-half of the excluded region, the loop number density is dominated by KK emission. The loop number density and $\omega_{cas}$ then 
\emph{decrease} when $\mu$ decreases, so that sufficiently small tensions are not constrained either.

The production and decay of KK modes at earlier times are constrained by their effects on the light elements synthetised by BBN. These constraints are usually derived for a background of long-lived relic particles at rest, which decay when the cosmic time is of the order of their lifetime, see~\cite{kawasaki} for a recent study. Two kinds of processes affecting the abundance of the light elements are distinguished, the radiative and hadronic ones. The radiative processes are due to the photon background produced by the decaying particles through electromagnetic cascade. When they are energetic enough, these photons produce $D$ and $^3 He$ by dissociation of $^4 He$, thus altering the successful predictions of standard 
BBN~\cite{lindley}. Similarly to the diffuse gamma-ray background, this constrains the total energy density injected into the cosmological medium. The injection of hadrons, on the other hand, leads to inter-conversions between neutrons and 
protons~\cite{inter} and hadro-dissociation of background nuclei~\cite{hadro}. The resulting constraints depend on both the number density and the energy of the injected hadrons. The constraints derived for a background of relatively light relic particles at rest cannot therefore be directly applied to our case of heavier and highly boosted KK modes produced by cusps. On the other hand, the constraints from radiative processes can be directly applied, because they depend only on the total energy that goes into the photon background produced by electromagnetic cascade. We will therefore focus on the latter. 

For long-lived particles with lifetime $\tau_X$, the bounds from radiative processes obtained in Ref.~\cite{kawasaki} (see Fig.~42 of that paper) read~\footnote{Constraints from CMB distortions~\cite{husilk} are typically milder, except for 
$\tau_X \sim 10^{12}$ sec where they become comparable.} $\rho_X / s \lesssim 10^{-14}$ GeV for $\tau_X$ ranging from about 
$10^8$ to $10^{12}$ sec. Here $\rho_X / s$ is the energy density of the long-lived particles in units of the entropy density before they decay. Contrary to long-lived relic particles decaying at time $t \sim \tau_X$, KK modes are produced by cusps and decay continuously. Since in the former case the decay occurs mainly in one Hubble time around 
$\tau_X$, we will impose the constraints of \cite{kawasaki} on the energy density injected by the KK modes decay 
in one Hubble time, $\Delta \rho_{inj} \approx t\,\dot{\rho}_{KK}$. We expect this approximation to be conservative because it neglects the cumulative effects that the continuous decay of the KK modes may have. Using (\ref{drhorad}) in the radiation-dominated era, this gives
\be
\label{rhoovers}
\frac{\Delta \rho_{inj}}{s} \, \approx \,  
\frac{10\,\Gamma_{KK}\,\zeta\,p^{-\beta}\,\sqrt{\alpha}}{\Gamma^2\,(G \mu)^{5/4}\,t}\,
\,\mathrm{Min}\left[1 \, , \, \frac{t^{2/3}}{t_{\mu}^{2/3}}\right]
\ee
where we used $s \sim 0.1 / (t^{3/2}\,G^{3/4})$ for the epoch under consideration. As discussed above, 
$\Delta \rho_{inj} / s$ should be less than $10^{-14}$ GeV for $t$ ranging from $10^8$ to $10^{12}$ sec. It is clear from (\ref{rhoovers}) that the strongest constraint comes from the smaller time in this interval. Ref.~\cite{kawasaki} also obtains similar constraints for long-lived particles decaying at earlier times, 
$1\,\mathrm{sec} \,\lesssim\, t \,\lesssim\, 10^7\,\mathrm{sec}$. However, these constraints result from hadronic processes and are therefore not directly applicable to our case, as discussed above. We note however that hadronic processes could lead to stronger constraints on cosmic super-strings.   

\begin{figure}[htb]
\begin{center}
\includegraphics[width=10cm]{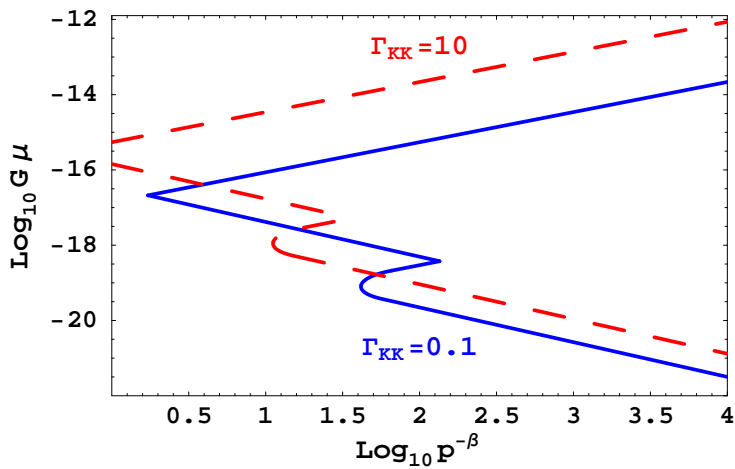}
\end{center}
\vspace*{-5mm}
\caption{Same as Fig.~\ref{Ga1} but for $\Gamma_{KK} = 0.1$ (blue, plain lines) and $\Gamma_{KK} = 10$ (red, dashed lines).}
\label{Gad}
\end{figure}

The constraint on cosmic super-strings in the $(p^{-\beta}, G \mu)$-plane from the photo-dissociation of light elements is shown in Fig.~\ref{Ga1} for the same values of the parameters as before. The region inside the two straight lines (in red) in the upper-half of the plot is excluded by BBN. The qualitative behavior of the constraint when $\mu$ and $p$ vary is the same as for the diffuse gamma-ray background. However, BBN constrains larger values of the string tension. In 
Fig.~\ref{Gad}, we also show the combined constraints from BBN and the diffuse gamma-ray background for other values of the parameter $\Gamma_{KK}$.

It is clear from Figs.~\ref{Ga1} and \ref{Gad} that KK emission is constrained mainly for relatively small reconnection probabilities $p$ or large values of the parameter $\Gamma_{KK}$. For cusp annihilation on standard Abelian-Higgs cosmic strings, with $p = 1$ and $\Gamma_{part} \sim 1$, only a very small range of string tensions around $G \mu \sim 10^{-16}$ might be constrained, although a more careful analysis would be required in that case. For cosmic super-strings, the reconnection probability depends~\cite{psuper} on the string coupling $g_s$, on the kind of strings that are present or dominate the loop number density, and on a volume suppression factor that depends on the compactification details. 
This latter factor comes from quantum fluctuations of the strings in the internal space. On the other hand, the parameter 
$\Gamma_{KK}$ defined in Eq.~(\ref{Gammakk}) depends on $g_s$, on the compactification details through the parameter 
$\kappa_E$ that encodes the effects of the KK spectrum, and on the kind of strings that are present or dominate the loop number density ($\chi = 0$ for F-strings and $\chi = 1$ for D-strings). The constraints also depend on the average number of cusps per loop oscillation period $c$ and on the way that the loop number density varies with $p$, through the parameter 
$\beta$ defined in (\ref{drholoop}). The values $c = 1$ and $\beta = 1$ are often expected to be good approximations, so we now further discuss the constraints in that case.

\begin{figure}[htb]
\begin{center}
\includegraphics[width=10cm]{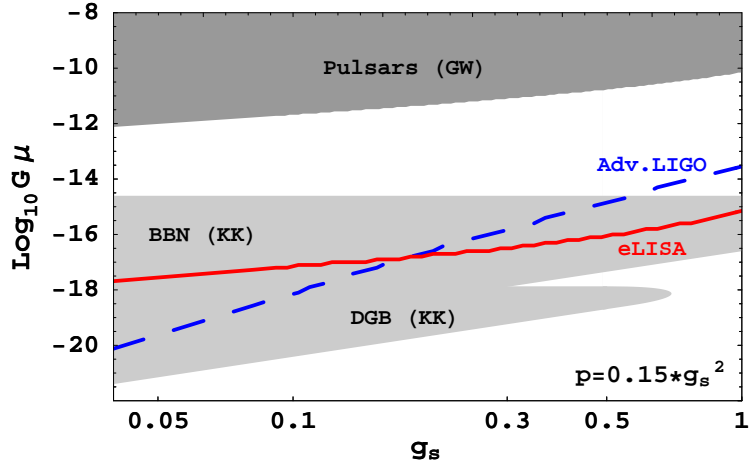}
\end{center}
\vspace*{-5mm}
\caption{BBN and DGB constraints from KK emission for F-strings in a KS throat, with $\zeta = 10$, $\alpha = 0.1$, 
$\Gamma = 50$, $c = \beta = 1$ and $\tilde{R} / \sqrt{\alpha'} \sim 10$ (see the main text for details). We also show for comparison the regions of the parameter space for which the stochastic GW bacground produced by the loops is accessible to different GW experiments, calculated from \cite{elisa}. The lower shaded region is excluded by either BBN (in its upper part) or DGB (in its lower part) due to KK emission. The upper shaded region is excluded by current pulsar timing observations due to GW emission. Finally, the region above the (blue) dashed line can be probed by Advanced LIGO, while the region above the (red) plain line is accessible to the ESA space interferometer eLISA~\cite{pau}.}
\label{bbndgbgw}
\end{figure}

The parameters $\Gamma_{KK}$ and $p$ are in general not independent, since they both depend on the string coupling 
$g_s$ and on the compactification details, in particular on the compactification scale $R / \sqrt{\alpha'}$. To illustrate this, we now focus on the most studied example of throat geometry, which is based on the Klebanov-Strassler (KS) solution discussed in sub-section~\ref{throat}. We also assume that only F-strings are present or dominate the loop number density. When different kinds of strings are present and interact, we expect the F-strings to dominate the loop number density, because they typically have the smallest reconnection probabilities~\cite{psuper}. The reconnection probabilities for cosmic super-strings in a KS throat depend on the ratio $\tilde{R} / \sqrt{\alpha'}$ given in Eq.~(\ref{gsM}). They were studied in detail in \cite{psuper}. The results of that paper show that, except for 
$1 < \tilde{R} / \sqrt{\alpha'} \lsim 4$, $p$ depends only logarithmically on $\tilde{R} / \sqrt{\alpha'}$ for F-strings, with $p \approx 0.15 g_s^2$ for $\tilde{R} / \sqrt{\alpha'} \sim 10$. As discussed below Eq.~(\ref{sumn}), $\kappa_E$ depends also weakly on $\tilde{R} / \sqrt{\alpha'}$ in that case, with $\kappa_E \approx 10$ for a range of values around 
$\tilde{R} / \sqrt{\alpha'} \sim 10$. For $c = 1$, Eq.~(\ref{Gammakk}) then gives $\Gamma_{KK} \approx 10 g_s^2$ for F-strings ($\chi = 0$). 
The BBN and DGB constraints from KK emission for these values of $\Gamma_{KK}$ and $p$ are shown in Fig.~\ref{bbndgbgw} in the $(g_s, G \mu)$-plane. These constraints exclude the lower shaded region on the figure. 

For comparison, we also show in Fig.~\ref{bbndgbgw} the regions of the parameter space for which the stochastic GW background produced by the loops is accessible to current and future GW experiments, as discussed in \cite{elisa}. The upper shaded region on the figure is excluded by current pulsar timing observations for these values of the parameters. The region above the (blue) dashed line can be probed by Advanced LIGO, while the region above the (red) plain line is accessible to the ESA space interferometer eLISA~\cite{pau}. Note that the effect of particle production on the loop number density was not taken into account in the calculation of the GW background in \cite{elisa}. The smallest values of the string tension that are accessible to eLISA in Fig.~\ref{bbndgbgw} correspond to GW that were produced at redshifts 
$z \sim 1$. These should not be modified by particle production. On the other hand, particle production could affect the smallest values of the string tension that are accessible to Advanced LIGO. This however would not change our conclusion about the complementarity between KK and GW emission in probing the parameter space of cosmic super-strings.

We see from Fig.~\ref{bbndgbgw} that the combination of the current constraints coming from GW emission (pulsars) and KK emission (BBN and DGB) leads already to stringent limits on $G \mu$. In particular, for $g_s \sim 0.05$, the string tension is restricted to the ranges $10^{-15} \lsim G \mu \lsim 10^{-12}$ or $G \mu \lsim 10^{-21}$. Furthermore, the whole region between the two shaded areas in Fig.~\ref{bbndgbgw}, which is currently unconstrained, will be accessible to either Advanced LIGO or eLISA. If no cosmological GW background is found by these experiments, the lower value of $G \mu$ that is constrained by KK emission would then set the strongest upper bound on the string tension. This upper bound reads 
$G \mu \lsim 2.5 \times 10^{-17} \, g_s^{32/13}$ for the BBN constraint ($g_s \sim 0.3 - 1$ in Fig.~\ref{bbndgbgw}) and 
$G \mu \lsim 10^{-18} \, g_s^{32/13}$ for the DGB constraint ($g_s \lsim 0.3$ in Fig.~\ref{bbndgbgw}). 

The best outcome, however, is of course that these experiments do observe a stochastic GW background. Unfortunately, the predictions for the GW spectrum from cosmic super-strings are highly degenerate in the $(p, G \mu)$-plane. Therefore, even in the most optimistic case where we can gain confidence in the fact that the observed background is produced by cosmic strings, it will not be possible to determine the fundamental parameters. The BBN and DGB constraints will then allow to discriminate between a wide range of possibilities, because they rule out a significant part of the cosmic super-string parameter space that would otherwise be accessible to Advanced LIGO or eLISA, as shown in Fig.~\ref{bbndgbgw}. 

Finally, it is easily checked that the friction-dominated epoch does not affect the above constraints from KK emission. As discussed in sub-section~\ref{friction}, the friction-dominated epoch may affect the rate of KK energy density emission when the loops decay mainly by KK emission, i.e. when $t < t_{\mu}$ and when $\dot{\rho}_{KK}$ is dominated by the loops with length $L = L_{part}(t)$. These loops are only present at times $t > t_*$, where $t_*$ is given in (\ref{tstar}). We must check that this is satisfied at $t \sim 10^8$ sec for the BBN constraint and at $t \sim t_0$ for the DGB one. This requires respectively $G \mu \gtrsim 10^{-19}$ and $G \mu \gtrsim 10^{-22}$ for $\Gamma_{part} \sim 1$, $\alpha \sim 0.1$ and the value of $t_d$ given in (\ref{tdFT}). Thus the constraints from KK emission are not affected by the friction-dominated epoch for the whole range of string tensions shown in Figs.~\ref{Ga1} to \ref{bbndgbgw}. For the smaller value of $t_d$ given in (\ref{tdST}), an even wider range of string tensions is constrained for smaller reconnection probabilities.

\section{Conclusion}
\label{conclu}

Cosmic super-strings surviving until the present epoch may open new observational windows into string theory physics. 
When they are produced at the end of brane inflation, and in models where the CMB anisotropies come only from the quantum fluctuations of a single inflaton field, their tension is expected to be in the range 
$10^{-13} \lesssim G \mu \lesssim 10^{-7}$ or so. The gravitational effects of such cosmic strings can be observable in the near future, in particular with upcoming gravitational wave experiments. In general, however, cosmic super-strings may be much lighter. As discussed in the Introduction, this occurs in particular when they are produced in models of brane inflation at lower energy scales, or alternatively at Hagedorn phase transitions after inflation. Since their gravitational effects are much weaker, it is important to look for other possible signatures of such light cosmic strings, in particular through the production of particles. 

In this paper, we studied the production of KK modes by cusps on loops of cosmic F- and D-strings. The masses of the KK modes and their couplings to the cosmic strings are set by the geometry of the internal space, which also determines the effective string tension $\mu$ percieved in four dimensions. For cosmic strings localized at the bottom of a warped throat, relatively few KK modes are lighter than the string scale, but they are strongly coupled to the cosmic strings. In the case of a flat internal space with large volume, the KK modes are only gravitationally coupled, but many of them are very light. We found that the total energy emitted by cusps in KK modes is comparable in both cases. It is also comparable (up to a numerical factor of order unity that depends on the string coupling $g_s$ and on the compactification details) to the energy emitted in scalar and gauge fields by cusp annihilation on standard Abelian-Higgs cosmic strings. However, we calculated the perturbative production of KK modes within the regime of validity of the effective Nambu-Goto description, while the process of cusp annihilation is non-perturbative and occurs beyond this regime. We also found that the total number of KK modes emitted by cusps is strongly enhanced in the case of a single large extra dimension. 

We then addressed cosmological consequences of KK emission by cosmic super-strings. We focused on the case where the loops are produced from the string network with a large initial size ($L_i \sim 0.1\,t$), as indicated by the most recent simulations~\cite{simul1, simul2} (see the discussion above Eq.~(\ref{Li})). We showed that KK emission is constrained by observations of the diffuse gamma-ray background and by the photo-dissociation of the light elements produced by BBN, as shown in Figs.~\ref{Ga1} to \ref{bbndgbgw}. These constraints rule out regions of the parameter space of cosmic super-strings that are complementary to the regions that can be probed from the strings' gravitational effects. In particular, the combination of the current constraints coming from GW and KK emissions leads already to stringent limits on the string tension, see Figs.~\ref{bbndgbgw}. Furthermore, the regions of the parameter space that are ruled out by KK emission both extend and overlap with the regions that lead to a stochastic GW background accessible to GW detectors such as Advanced LIGO and the ESA space interferometer eLISA~\cite{elisa}. The constraints from KK emission will then either extend the range of excluded parameters if no GW from cosmic strings are observed, or help to discriminate between a wide range of possibilities in the case of a detection.  

On the string theory side, the constraints from KK emission are sensitive to the string coupling $g_s$ and to the geometry of the internal space. In Fig.~\ref{bbndgbgw}, we specified the constraints for the most developed model based on the so-called Klebanov-Strassler throat. If other string theory models emerge, our calculation can be extended to other geometries of the internal space. This essentially requires to calculate the reconnection probabilities $p$ and the coefficient 
$\kappa_E$ defined in Eq.~(\ref{kappaE}) that encodes the effect of the KK spectrum. The constraints depend also on the typical size of the loops when they are produced, on the way the loop number density varies with $p$, and to a lesser degree on the average number of cusps per loop oscillation. These properties of cosmic strings are still uncertain and can hopefully be determined by the next generation of simulations. We expect however our main conclusion - the complementarity between KK and GW emissions in probing the parameter space of cosmic super-strings - to be qualitatively robust. This may ultimately allow to constrain or determine fundamental parameters of string theory, such as the string coupling $g_s$ and the compactification details.

We have only considered KK modes that correspond to spin-$2$ fields in four dimensions, because these are the most generic ones, and we calculated their production within the regime of validity of the effective Nambu-Goto description. In general, however, other KK modes may be produced too. Furthermore, beyond the Nambu-Goto description, one may expect cusps to emit string states, which would then decay into extra KK modes. It would be interesting to study this process in string theory, see \cite{skliros} for recent developments in this direction. If more KK modes are produced, this may widen the ranges of string tensions that are constrained by KK emission towards larger values.

That KK emission leads to constraints in a range of string tensions, as opposed to just a lower or an upper bound, comes from the following reason. For large string tensions, the loops decay relatively quickly by gravitational emission, so their number density is relatively small. As the tension decreases, the lifetime of the loops increases, and thus also their number density, leading to a larger energy density emitted in KK modes and therefore to stronger cosmological consequences. However, for sufficiently small values of the string tension, the lifetime of the loops starts being dominated by KK emission and becomes shorter. Further decreasing the tension then decreases the loop number density and the energy density emitted in KK modes, so that the constraints eventually disappear. It was therefore important for us to carefully take into account the effects of KK emission on the loop number density. These effects are important mainly for small string tensions or in the very early universe. We only presented here the modifications of the loop number density in the regimes of interest in this paper, leaving a more complete analysis and the discussion of other cosmological consequences for future work. 

We have followed the standard expectation that, except potentially in the very early universe, the cosmic super-strings are localized at a classically fixed position in the internal space~\cite{psuper, pol}. In Ref.~\cite{gregory}, it was shown that the GW signal emitted from cusps is reduced if the cosmic strings can move in a flat internal space. In that case, KK emission would be suppressed too. This would presumably also lead to very small reconnection probabilities. However, as emphasized in \cite{psuper}, the position of the strings in the internal space corresponds to worldsheet moduli that are not protected by any symmetry and should therefore be fixed at the minimum of their effective potential. In that case, there is no classical motion of the strings in the extra-dimensions, but the reconnection probability is reduced by the
quantum fluctuations of the strings in the internal space. 

It was also found in Ref.~\cite{avgoustidis} that, given initial conditions for the string motion and a warp factor generating an effective potential for their position, Hubble damping may not be efficient enough to localize the strings in the internal space after their production. We note however that other mechanisms of damping may be more efficient. As we have mentioned several times, KK modes are also abundantly produced when the cosmic strings form, either thermally or as 
the decay products of brane / anti-brane annihilation. Similarly to the friction-dominated epoch for field theory cosmic strings, we therefore expect the cosmic super-strings interacting with this thermal gas of KK modes to experience a friction force that can stabilize their position in the internal space and damp their motion in the non-compact dimensions. We provided an estimate for the timescale of this process in Eq.~(\ref{tdST}). The friction-dominated epoch for cosmic 
super-strings is clearly interesting and deserves further investigations.

To conclude, cosmic super-strings come generically together with light (compared to the string scale), and therefore cosmologically relevant, KK modes. They offer potential signatures of string cosmology. Other cosmological consequences of KK emission by cosmic super-strings remain to be studied, e.g. the production of cosmic rays or dangerous relics.


\section*{Acknowledgments}

It is a pleasure to thank Denis Allard, Pierre Bin\'etruy, Alejandro Bohe, Marc Lilley, Marco Peloso, Eray Sabancilar, Lorenzo Sorbo, Tanmay Vachaspati and especially Dani\`ele Steer for useful discussions and correspondence.


\section*{APPENDIX}

In this Appendix, we calculate in detail the amount of massive particles produced by a cusp in the stationary phase approximation. 

The interaction (\ref{int2}) leads to the equation of motion
\be
\left(\Box + m_{\bar{n}}^2\right)\,h^{\bar{n}}_{\mu \nu} = -\,\frac{\lambda_{\bar{n}}}{\sqrt{\mu}}\,T^\mathrm{TT}_{\mu\nu}
\ee
for each spin-$2$ field $h^{\bar{n}}_{\mu \nu}$, where $T^\mathrm{TT}_{\mu\nu}$ is the transverse and traceless part of the energy-momentum tensor $T_{\mu\nu}$ of the loop, 
$\partial^\mu T^\mathrm{TT}_{\mu\nu} = \eta^{\mu\nu} T^\mathrm{TT}_{\mu\nu} = 0$. The energy emitted in KK modes with mode numbers $\bar{n}$ by the source can be calculated as
\be
\label{defEn}
E_{\bar{n}} = \frac{\lambda_{\bar{n}}^2}{\mu}\,\int \frac{d^3\mathbf{k}}{(2 \pi)^3}\,\frac{1}{2}\,
T^{\mathrm{TT} \mu\nu}(\omega_k, \mathbf{k})\,T^{\mathrm{TT} *}_{\mu\nu}(\omega_k, \mathbf{k})
\ee
where the double Fourier transform $T^{\mathrm{TT}}_{\mu\nu}(\omega_k, \mathbf{k})$ is defined as in 
Eqs.~(\ref{FT}, \ref{4k}). For $k^\lambda\,k_\lambda = m_{\bar{n}}^2 \neq 0$, the transverse-traceless part of the energy-momentum tensor in Fourier space can be taken with the projectors
\be
\label{TTpart}
T^\mathrm{TT}_{\mu\nu}(\omega_k, \mathbf{k}) = \mathcal{O}_{\mu \nu \rho \sigma}\,
T^{\rho \sigma}(\omega_k, \mathbf{k}) = 
\left(P_{\mu\rho}\,P_{\nu\sigma} - \frac{1}{3}\,P_{\mu\nu}\,P_{\rho\sigma}\right)\,
T^{\rho \sigma}(\omega_k, \mathbf{k})
\ee 
where
\be
\label{Pmunu}
P_{\mu\nu} = \eta_{\mu\nu} - \frac{k_\mu\,k_\nu}{m_{\bar{n}}^2} \, .
\ee
Using the property $\mathcal{O}_{\mu \nu \rho \sigma}\,\mathcal{O}^{\mu \nu}_{\hspace*{0.35cm}\kappa \lambda} = 
\mathcal{O}_{\rho \sigma \kappa \lambda}$, and the fact that $T_{\mu \nu}(\omega_k, \mathbf{k})$ is already transverse 
(i.e. $k^\mu \, T_{\mu\nu} = 0$, since $T_{\mu\nu}$ is conserved), Eq.~(\ref{Em}) follows from 
Eqs.~(\ref{defEn}, \ref{TTpart}).

The dynamics of the loop resulting from the Nambu-Goto action (\ref{NGS}) is conveniently studied in the conformally flat gauge
\be
\label{gaugecons}
\partial_{\tau}X^{\lambda} \partial_{\sigma}X_{\lambda} \, = \, 
\partial_{\tau}X^{\lambda} \partial_{\tau}X_{\lambda} + \partial_{\sigma}X^{\lambda} \partial_{\sigma}X_{\lambda}
\, = \, 0
\ee
where $\gamma_{\alpha\beta} = \sqrt{-\gamma}\,\eta_{\alpha\beta}$. The solution of the two-dimensional wave equation for the string reads
\be
\label{Xpm}
X^\mu(\tau, \sigma) = \frac{1}{2}\,\left[X_+^\mu(\sigma_+) + X_-^\mu(\sigma_-)\right] \hspace*{0.5cm} \mbox{where} 
\hspace*{0.5cm} \sigma_{\pm} = \tau \pm \sigma \, .
\ee
It follows from the the gauge constraints (\ref{gaugecons}) that $\dot{X}_+^\mu$ and $\dot{X}_-^\mu$ are null vectors, 
$\dot{X}_+ . \dot{X}_+ = \dot{X}_- . \dot{X}_- = 0$, where a dot on $X_+^\mu(\sigma_+)$ or $X_-^\mu(\sigma_-)$ denotes the derivative with respect to the corresponding unique variable, $\sigma_+$ or $\sigma_-$ respectively. Choosing furthermore the time gauge where $\tau$ coincides with the Lorentz time in the center-of-mass frame of the loop, i.e. $t = X^0(\tau, \sigma) = \tau$, we have 
$X^0_{\pm}(\sigma_\pm) = \sigma_{\pm}$. The dynamics of the loop is then described by the two $3$-vectors $\mathbf{X}_+$ and $\mathbf{X}_-$. These are periodic functions of $\sigma_\pm$ of period $L$, where $L$ is the invariant length of the loop, and their derivative has unit norm, $|\dot{\mathbf{X}}_+| = |\dot{\mathbf{X}}_-| = 1$. 

We can choose the coordinates in Eq.~(\ref{Xpm}) such that the cusp occurs at $\sigma_\pm = 0$ and $X^\mu = 0$. At the cusp itself, we have
\be
l^\mu \equiv \left(1, \mathbf{n}\right) \equiv \dot{X}_+^\mu(0) = \dot{X}_-^\mu(0)
\ee
where $\mathbf{n}^2 = 1$. In the vicinity of the cusp ($\sigma_{\pm} \ll L$), we have
\be
\label{closecusp}
X_\pm^\mu(\sigma_\pm) \simeq l^\mu\,\sigma_\pm + \frac{1}{2}\,\ddot{X}_\pm^\mu(0)\,\sigma_\pm^2 + 
\frac{1}{6}\,X_\pm^{(3)\mu}(0)\,\sigma_\pm^3 
\ee
and the gauge constraints (\ref{gaugecons}) impose
\be
\label{gaugecons2}
l . l = l . \ddot{X}_\pm(0) = 0 \hspace*{0.5cm} , \hspace*{0.5cm}
l . X_\pm^{(3)}(0) = - \ddot{X}_{\pm}^2(0) \, .
\ee

It is convenient to introduce the vector
\be
\label{deltamu}
d^\mu \equiv l^\mu - \frac{k^\mu}{\omega_k} = (0, \mathbf{d}) 
\hspace*{0.5cm} \mbox{where} \hspace*{0.5cm} \mathbf{d} = \mathbf{n} - \frac{\mathbf{k}}{\omega_k} \, .
\ee
As discussed below Eq.~(\ref{stationary}), the phases in the integrals (\ref{TmunuDV}) are smaller than unity for 
$m_{\bar{n}} \ll k \simeq \omega_k$ and $\theta \ll 1$, where $\theta$ is the angle between $\mathbf{k}$ and $\mathbf{n}$. We then have $\mathbf{d}^2 \simeq \theta^2 + m_{\bar{n}}^4 / (4 k^4) \ll 1$. The phases in (\ref{TmunuDV}) can be calculated from Eqs.~(\ref{closecusp}), (\ref{gaugecons2}) and (\ref{deltamu}) as
\be
\frac{1}{2}\,k_\lambda \,X_\pm^\lambda \, \simeq \, \frac{\omega_k}{2}\,\mathbf{d}\,\mathbf{n}\,\sigma_\pm + 
\frac{\omega_k}{12}\,\ddot{X}_\pm^2(0) \, \sigma_\pm^3
\ee
where we have neglected terms of higher order in $\sigma_\pm / L$. Using (\ref{ddotXL}) and the expression for $\mathbf{d}$ above, we then get Eq.~(\ref{phases}).

From (\ref{TmunuDV}), the integrand in Eq.~(\ref{Em}) can be calculated as
\be
\label{TmunuI}
T^{\mu\nu}(\omega_k, \mathbf{k})\,T_{\mu\nu}^{*}(\omega_k, \mathbf{k}) - 
\frac{1}{3}\,|T^{\lambda}_{\lambda}(\omega_k, \mathbf{k})|^2 \, = \, 
\frac{\mu^2}{8}\,\left(|I_+|^2\,|I_-|^2 + |I_+ . I_-^*|^2 - \frac{2}{3}\,|I_+ . I_-|^2\right)
\ee
where we use the notations $|I_+|^2 = I_+^\lambda \, I_{+ \, \lambda}^*$, 
$I_+ . I_-^* = I_+^\lambda \, I_{- \, \lambda}^*$, etc. Using the expansion (\ref{closecusp}) in (\ref{TmunuDV}), we have
\be
I_{\pm}^{\mu} \approx A\,l^\mu + B\,\ddot{X}_\pm^\mu(0) + ...
\ee
where we have defined
\be
\label{defAB}
A = \int d\sigma_\pm\,e^{\frac{i}{2}\,k_\lambda\,X_\pm^\lambda} \hspace*{0.5cm} \mbox{ and }  
\hspace*{0.5cm} B = \int d\sigma_\pm\,\sigma_{\pm}\,e^{\frac{i}{2}\,k_\lambda\,X_\pm^\lambda} \, .
\ee
Inserting into (\ref{TmunuI}) and using (\ref{ddotXL}) gives
\be
\label{TTcal}
T^{\mu\nu}(\omega_k, \mathbf{k})\,T_{\mu\nu}^{*}(\omega_k, \mathbf{k}) - 
\frac{1}{3}\,|T^{\lambda}_{\lambda}(\omega_k, \mathbf{k})|^2 \, \approx \, 
\frac{\mu^2}{8}\,\left(\frac{2 \pi}{L}\right)^4\,|B|^4
\ee
up to a factor of order unity that depends on the shape of the cusp. The leading terms in $A^2$ and $A\,B$ in $I_\pm^2$ canceled because of the gauge constraints (\ref{gaugecons2}).

The coefficient $B$ defined in (\ref{defAB}) can be calculated analytically. Using Eq.~(\ref{phases}) for the phases and defining
\be
\label{defu}
u = \left(\theta^2 + \frac{m^2_{\bar{n}}}{k^2}\right)^{-1/2} \, \frac{2 \pi}{L} \, \sigma_\pm
\ee
and
\be 
\label{defv}
v = \frac{4}{3 \pi}\,\left(\frac{\theta^2}{\theta_c^2} + \frac{k_c^{4/3}}{k^{4/3}}\right)^{3/2} \, ,
\ee
with $k_c$ and $\theta_c$ given in (\ref{kc}, \ref{thetac}), we have
\bea
\label{Bcal}
B &=&  \left(\frac{L}{2 \pi}\right)^2 \, \left(\theta^2 + \frac{m^2_{\bar{n}}}{k^2}\right) \, 
\int_{-\infty}^{+\infty} du \, u \, e^{i\,\frac{3}{2}\,v\,(u + \frac{u^3}{3})} \nonumber \\ 
&=& \left(\frac{L}{2 \pi}\right)^2 \, \left(\theta^2 + \frac{m^2_{\bar{n}}}{k^2}\right) \, 
\frac{2\,i}{\sqrt{3}} \, K_{2/3}(v)
\eea
where $K_{\nu}(z)$ is the modified Bessel function of the second kind~\cite{abramowitz}. In the stationary phase approximation, the first factor in the RHS of (\ref{defu}) is much bigger than unity and the limits of the integral above are extended to infinity. 

\begin{figure}[htb]
\begin{center}
\begin{tabular}{cc}
\includegraphics[width=8cm]{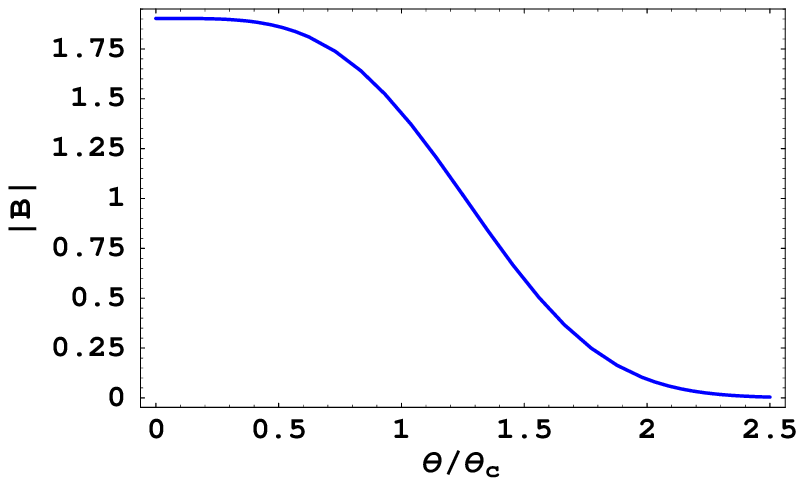} \hspace*{1cm}
\includegraphics[width=8cm]{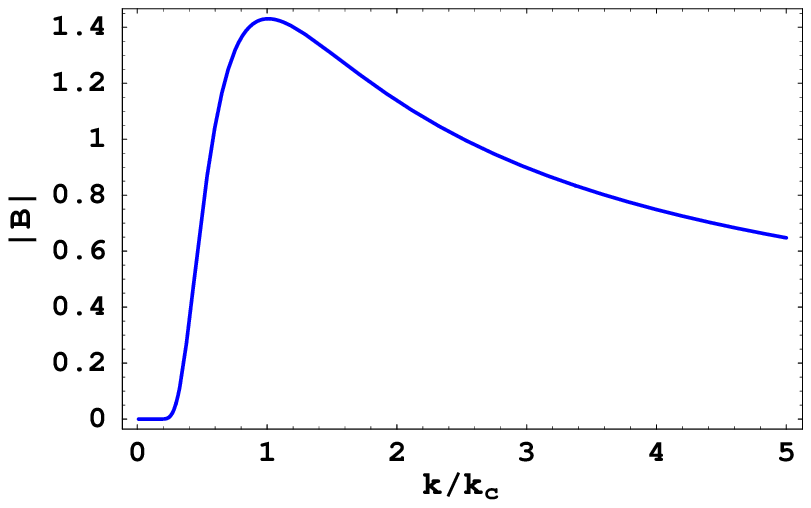}
\end{tabular}
\caption{Variation of $|B|$ (arbitrarily normalized) with $\theta / \theta_c$ and $k / k_c$. The left panel is for 
$m_{\bar{n}} = 0$ or $k \gg k_c$. The right panel is for $\theta = 0$.}
\label{Bbehav}
\end{center}
\end{figure} 

The behavior of $|B|$ as a function of $\theta$ and $k$ is illustrated in Fig.~\ref{Bbehav}. At large argument $v \gg 1$, we have $K_{2/3}(v) \propto e^{-v} / \sqrt{v}$. Thus $B$ is exponentially suppressed for $\theta \gg \theta_c$ or 
$k \ll k_c$, as expected. At small argument $v \ll 1$, we have $K_{2/3}(v) \propto v^{-2/3}$. Therefore, for $k \gg k_c$ and a fixed ratio $\theta / \theta_c$, $B$ varies as $k^{-2/3}$. When $k$ increases above $k_c$, the beaming angle 
(\ref{thetac}) decreases, which further suppresses the contribution of these modes. When $k$ decreases below $k_c$, 
the beaming angle increases but this is quickly compensated by the exponential decrease of $B$ for $k \ll k_c$. 

Inserting (\ref{TTcal}) with (\ref{Bcal}) into (\ref{Em}), and with a suitable change of variables for the integral over 
$d\mathbf{k}$, we obtain Eq.~(\ref{En}) with
\be
C = \frac{32}{3 \pi^6}\,\int_0^{+ \infty} dy\,y^{-3/4}\,\int_0^{\frac{m L}{4 \sqrt{y}}} dx \, (x + y)^4\,
K_{3/2}^4\left(\frac{4}{3 \pi}\,(x + y)^{3/2}\right) 
\ee
which does not depend on $m_{\bar{n}} L$ for $m_{\bar{n}} L \gg 1$. Performing the integral numerically gives 
$C \simeq 0.4$. The coefficient appearing in Eq.~(\ref{Nn}) for the number of produced KK modes is obtained similarly as
\be
D = \frac{128}{3 \pi^6}\,\int_0^{+ \infty} dy\,\int_0^{\frac{m L}{4 \sqrt{y}}} dx \, (x + y)^4\,
K_{3/2}^4\left(\frac{4}{3 \pi}\,(x + y)^{3/2}\right)
\ee
which gives $D \simeq 0.3$.



\end{document}